\documentclass{article}
\usepackage{amsmath,amsthm,amsfonts,amssymb}
\usepackage[pdftex,pdfborder={0 0 0}]{hyperref}

\usepackage{fullpage}
\usepackage[noblocks]{authblk}
\usepackage{caption}
\usepackage{subcaption}
\usepackage{tikz}
\usepackage{float}
\usepackage{enumerate}
\usepackage{multirow}
\usepackage{adjustbox}
\usepackage{soul}
\usepackage{tabularx}% http://ctan.org/pkg/tabularx
\usepackage{booktabs}% http://ctan.org/pkg/booktabs
\usepackage[title]{appendix}
\newcolumntype{Y}{>{\raggedleft\arraybackslash}X}% raggedleft column X
\usepackage{graphicx} % Required for inserting images

\title{Harvests and Hooky in the Hills: Crop Yield Variability and Gendered School Enrollment in Rwanda}
\date{October 2025}

\begin{document}

\author[1]{Maxwell Fogler\thanks{Maxwell.Fogler@colorado.edu}}

\affil[1]{University of Colorado, Boulder, CO, USA}

\date{September 2025}

\maketitle

\begin{abstract}
    This paper investigates the trade-off that households in agrarian economies face between immediate production needs and long-term human capital investment. We ask how exogenous agricultural productivity shocks affect primary and secondary school enrollment in Rwanda, a country characterized by a heavy reliance on rain-fed agriculture alongside ambitious development goals. Using a district-level panel dataset for the years 2010-2021, we employ a two-stage least squares (2SLS) instrumental variable strategy. Plausibly exogenous variation in annual rainfall is used to instrument for a satellite-derived measure of vegetation health and agricultural productivity, the Normalized Difference Vegetation Index (NDVI), allowing for a causal interpretation of the results. For primary education, where direct costs are low, enrollment is countercyclical: positive productivity shocks are associated with lower enrollment. Notably, boys' primary enrollment is found to be significantly more elastic to these shocks than girls' enrollment. Conversely, for secondary education, which entails additional financial outlays, enrollment is strongly procyclical. Positive productivity shocks lead to significant increases in enrollment. Further, a sustained positive shock is associated with a subsequent decline in female secondary enrollment. The results challenge previous regional findings supporting the ``girls as a buffer'' hypothesis and investigate the dynamic and gendered responses to the persistence of economic shocks.
\end{abstract}

%\noindent JEL: D1, J1, J2, O1, Q1.\\

\footnotetext[1]{Special thanks to Dr. Shawn Swanson, Dr. Terra McKinnish, and the Ministry of Education of Rwanda. All errors are my own.}

\clearpage

\section{Introduction}
How households in agrarian economies navigate the trade-off between immediate production needs and long-term human capital investment, particularly when faced with income volatility, is a central question to development economics. The decisions households make in response to economic shocks can have lasting consequences for intergenerational mobility and national growth. The issue is further complicated by gender dynamics; intra-household allocation decisions may not affect sons and daughters equally, with significant implications for gender equality and economic efficiency (Jayachandran, 2014). This paper investigates these questions in the context of Rwanda, a country where a heavy reliance on rain-fed agriculture and exposure to climate-related risk coexist with a strong national commitment to developing a knowledge-based economy. This setting provides a compelling backdrop to examine how households manage agricultural productivity shocks and what these shocks reveal about the determinants of educational investment for boys and girls at both the primary and secondary levels.

Rwanda specifically serves as an interesting case study in how policy intervention can shape how households balance the tradeoff between the opportunity cost of education and labor. Released in 2020, the Vision 2050 document acts as a strategic policy plan which details the process through which Rwanda aims to become a middle-income country by the year 2050. A central tenet of the national development model is transitioning from primarily agrarian production to a knowledge-based economy. Emphasizing the importance of education, Rwandan legal frameworks assert that childhood schooling is a fundamental right which should not be dependent on household financial well-being (MINECOFIN, 2020).

Beginning in 2003, public sector school fees were abolished and funding for schools was provided through grants by the Rwandan Ministry of Finance (MINECOFIN) each year based on the number of enrolled pupils. According to the spending guidelines, 50\% of these funds should be used “for the functioning of the school,” which includes textbooks and other classroom materials, 35\% on infrastructure repair and sanitation, and 15\% on teacher capacity and training. Surveys found that 89\% of parents, 87.3\% of pupils, and 80.9\% of teachers were satisfied or very satisfied with how the grants were spent by their schools (Transparency International Rwanda, 2012).

Though nominal fee-free schooling has reduced the costs of schooling, it is important to note that it has not totally eliminated them. Students and parents alike identify external costs that are important for a cohesive social and educational experience - for example, uniforms are not covered but still required for attendance at many primary and secondary schools. For girls, sanitary pads add an additional cost and the lack thereof can result in substantial time missed from the classroom; though the government has made efforts to increase the availability of free hygiene products, there are still gaps in access, particularly in rural areas. Lastly, extracurricular costs, such as academic coaching for national examinations, can further exacerbate differences in the academic experience and outcome between high- and low-income households (Williams, 2025).

A household's decision to enroll a child in school in the face of an economic shock, particularly for low-income agrarian households, is governed by the interplay of two opposing forces: an income effect and a substitution effect. A positive productivity shock, such as a favorable harvest, increases household income, which may enable a family to better afford the direct and indirect costs of schooling (the income effect). Simultaneously, the same shock increases the marginal product of labor in agriculture, raising the opportunity cost of a child's time in school (Bai, 2020).

This framework raises a critical question: which effect prevails in a low-income but rapidly developing country like Rwanda, and does the answer differ for primary versus secondary education, where the economic costs of schooling diverge dramatically?

A robust collection of research has been published on the effects of income shocks on childhood health and performance.  A review of empirical literature by Ferreira et al.  (2008) on the subject finds that:

\begin{quote}
        \textit{In richer countries, like the United States, child health and education outcomes are counter-cyclical: they improve during recessions.  In poorer countries, mostly in Africa and low-income Asia, the outcomes are procyclical: infant mortality rises and school enrollment and nutrition fall during recessions.  In the middle-income countries of Latin America, the picture is more nuanced: health outcomes are generally procyclical and education outcomes counter-cyclical.}
\end{quote}

This topic is of high practical importance because, as previous long-term panels suggest, adverse experiences in early childhood result in worse outcomes in adulthood.  A study of Jamaican children whose height-for-age were two standard deviations or more below that of the general population were more likely to develop cognitive deficits, perform worse in school, and have poorer emotional and behavioral outcomes (Walker et al., 2007).  Similar results have been established in many other countries across the development spectrum (Hoddinott et al., 2008, Case and Paxson, 2008).

Despite the broad consensus that early childhood experiences significantly impact developmental outcomes, “there is a surprising dearth of population-based multinational data on the diverse experiences and conditions that promote or thwart child well-being” (Bornstein et al., 2012). The heterogeneity of production and lifestyle can make it difficult to establish general trends across diverse settings.

Jensen (2000) studies the effects of agricultural volatility on investments in children in Cote d'Ivoire.  Their findings suggest that investments in children decrease significantly during the presence of adverse agricultural conditions, with school enrollment rates declining by between one-third and one-half.

The literature on gendered responses to income shocks often posits a ``girls as a buffer'' hypothesis, where girls' time and consumption are used to shield the household, and particularly sons, from the negative consequences of a shock. Studies have shown that girls' well-being is often more sensitive to fluctuations in income and prices (Ojong, 2025).

In neighboring Uganda, for instance, Björkman-Nyqvist (2013) found that negative rainfall shocks had a significant negative effect on the enrollment of older girls, while boys' enrollment remained largely unaffected. This finding is consistent with a model where the value of girls' labor in domestic production is the primary margin of household adjustment. This established regional precedent provides a powerful point of comparison for our analysis, allowing us to investigate whether these gendered patterns of intra-household allocation are universal or are instead shaped by context-specific economic roles and cultural norms.

The primary empirical challenge in establishing a causal link between household economic conditions and educational investment is endogeneity; unobserved factors such as parental ability, policy changes, or preferences are likely correlated with both household income and children's schooling outcomes. To overcome this, this paper employs an identification strategy that exploits plausibly exogenous variation in agricultural productivity across Rwandan districts and over time. We use satellite-derived data on the Normalized Difference Vegetation Index (NDVI)—a direct measure of vegetation health and crop yield potential—as a proxy for agricultural shocks, instrumented by rainfall to account for potential measurement error or bias. This approach allows us to isolate the causal effect of productivity shocks on school enrollment.

This paper makes three distinct contributions to the literature on human capital, household economics, and gender. First, it provides robust causal evidence of a structural break in household decision-making between primary and secondary education. The results suggest that for primary schooling, the substitution effect is the dominant force, leading to countercyclical enrollment patterns. For secondary schooling, the income effect prevails, resulting in strongly procyclical enrollment. This finding empirically validates a bifurcated model of educational investment, where the binding constraint shifts from the opportunity cost of labor to financial capacity as a child progresses through the education system.

Second, the analysis presents a nuanced and contrasting view of gender dynamics at the primary level. Contrary to the findings in Uganda and the predictions of the ``girls as a buffer'' hypothesis, the evidence from Rwanda indicates that boys' primary school enrollment is more sensitive to agricultural productivity shocks than girls' enrollment. This suggests that the gendered division of labor and the elasticity of child labor substitution are highly context-specific.

Third, this study documents a complex temporal and gendered response to the lingering effects of shocks at the secondary level. While an positive shock initially boosts enrollment for both genders, a sustained period of agricultural prosperity is associated with a sharp subsequent decline in female enrollment, revealing how households' long-term strategic planning regarding education and labor markets adapts to changes in their economic environment. The remainder of this paper is organized as follows. Section 2 details the data sources and the empirical framework. Section 3 presents the primary results for primary and secondary enrollment. Section 4 provides a synthesis and discussion of the findings, and Section 5 concludes with policy implications.

\section{Methodology}
\subsection{Data Description}
This paper analyzes panel data where observations for all variables are spatially and temporally ordered by district and year respectively.  The primary results of this paper are obtained through a two-stage least squares instrumental variable fixed-effects model – which further considers the gender and education level of respondents when applicable.

This section will begin by describing the data sourcing, compilation, and limitations for both the explanatory and response variables.  Next, the overall trends are discussed with provided descriptive figures.  Lastly, this section ends with a description of the utilized instrumental variable derivation and regression equation.

%The Python notebooks and datasets used in this study have been included or, in cases where access requires government approval, described in the appendix.

\subsubsection{Crop Yield Data}
The primary explanatory variable of interest is crop yield.  Because the agricultural sector in Rwanda currently employs ~67\% of the labor force – with many being small-holder operations – and accounts for a significant portion of gross national production (~29\%), exogenous variation in Rwandan districts’ income over time caused by crop yield changes can be utilized to study the effects of negative household income shocks on youth educational attainment and enrollment (National Institute of Statistics of Rwanda, 2020).

All crop yield estimations were calculated using geospatial vegetation index values taken from satellite image data over the years 2001-2023.  The Normalized Difference Vegetation Index (NDVI) is a graphical indicator which remotely measures the green health and density of vegetation in a specific location.  Specifically, NDVI is calculated using reflectance values from spectrometric data at the red and near-infrared (NIR) bands as described in Eqn.  1.  These values will be between -1 and 1.  Values approaching 1 indicate temperate and highly dense areas of healthy green vegetation;  values close to 0 will generally correspond to barren rocky, snowy, or sandy areas; values close to -1 correspond to bodies of water.

\begin{equation}
    NDVI = \frac{NIR - Red}{NIR + Red}
\end{equation}

Specific interpretations of marginal changes in NDVI with respect to marginal changes are wide ranging and highly dependent on where and when the measurements were taken.  However, because many topological features of Rwanda are captured by NDVI – for example, Bucagu C.  et al.  showed NDVI to highly correlate with soil availability and fertility – combined with the fact that NDVI is linearly correlated with yield, absolute changes in NDVI year-over-year signify proportional absolute changes in yield.  I note this because direct district-level crop-yield figures are not available for all years in the study period and thus no direct NDVI-yield regressions could be performed.

Previous studies have investigated the use and accuracy of NDVI as a proxy for crop yield.  One such paper focused specifically on the ability of NDVI values to determine maize yield gaps in Rwanda and found that NDVI values were highly correlated with maize grain yield and can effectively detect granular spatial differences caused by Rwanda’s diverse set of microclimates (Bucagu, Charles, et al., 2020).  Overall, NDVI has been widely used in the research community as a reliable and robust indicator of crop yield – especially in cases where historical or low-level data is unreliable or unavailable.

The satellite imagery data were obtained from the Moderate Resolution Imaging Spectroradiometer (MODIS) MOD13 Series product.  A NASA project aboard the Terra and Aqua satellites, MODIS provides “consistent, spatial and temporal time series comparisons of global vegetations conditions that can be used to monitor the Earth’s terrestrial photosynthetic activity in support of phenologic, change detection, and biophysical interpretations” (Didan et al.  2019 – see source for description of MODIS characteristics).

\begin{figure}
    \centering
    \includegraphics[width=0.6\linewidth]{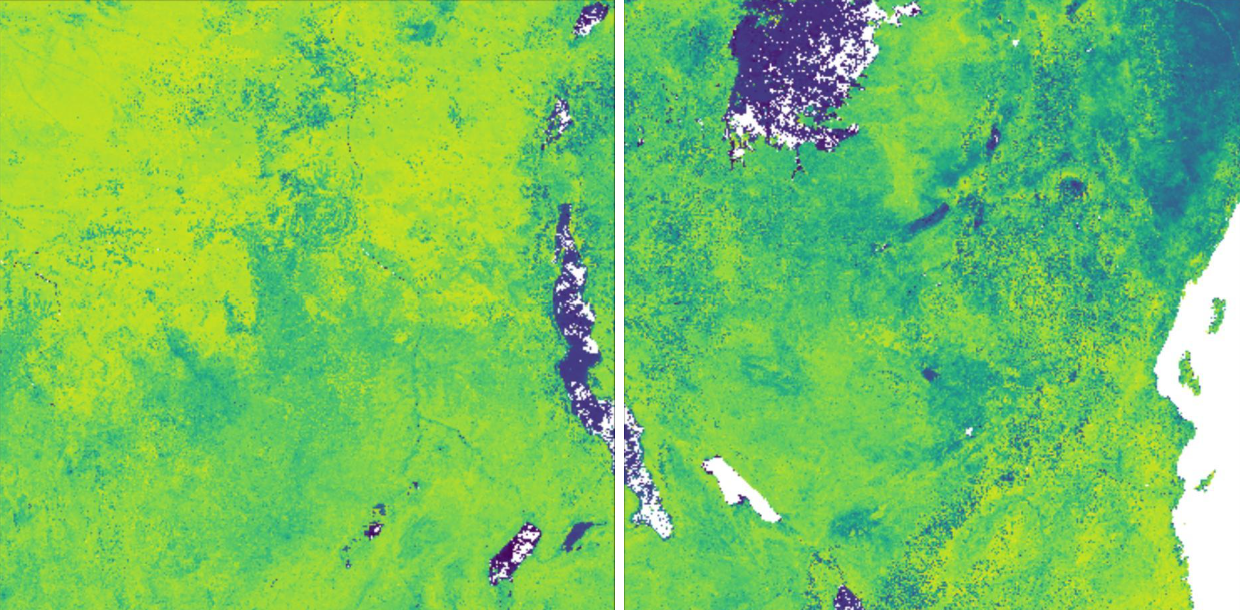}
    \caption{MODIS Tiles h20v09 and h21v09 Reprojected with NDVI Coloring}
    \label{fig:TilesReprojected}
\end{figure}

This paper utilizes satellite imagery from the years 2001 through 2023 which are taken in 16-day intervals and have a spatial resolution of 250m.  All data are separated into 460 ‘tiles’ which encapsulate approximately a 1,110km x 1,110km area and follow a sinusoidal projection.  Tiles h20v09 and h21v09 are used for this paper’s analysis as the former covers the western districts while the latter covers the eastern districts. The reprojected NDVI data are shown in Figure \ref{fig:TilesReprojected}.

\begin{figure}
    \centering
    \includegraphics[width=0.6\linewidth]{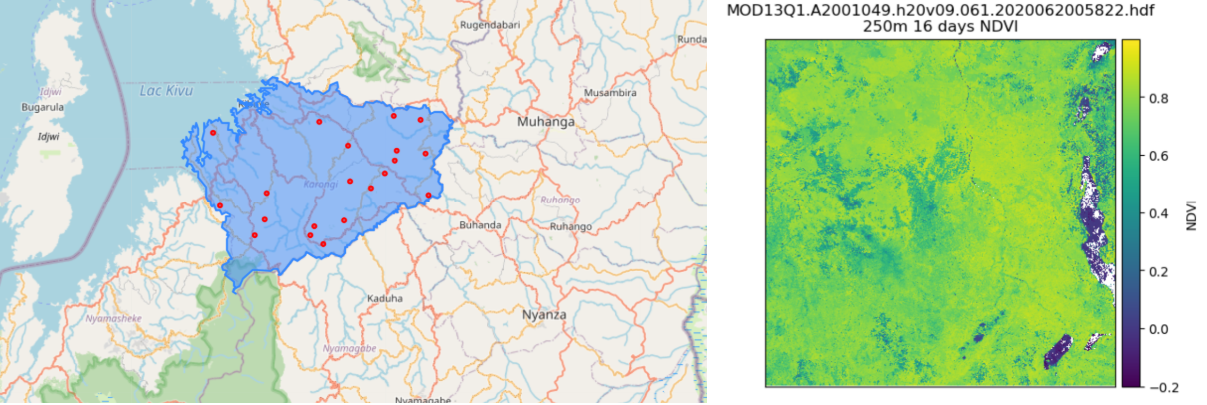}
    \caption{Example of 20 Sampled Points from Karongi and Corresponding MODIS Tile}
    \label{fig:KarongiSample}
\end{figure}

Once all MODIS data could be interpreted using the EPSG:4326 coordinate system, a shapefile of Rwandan districts was used to randomly sample a time invariant set of 1,000 pixels from each of the 30 districts which would be used to extract approximate district-wide mean NDVI for each of the 16-day periods for the years 2001-2023. A subset of sampled points are shown in Figure \ref{fig:KarongiSample}.

The limitations of this data include the variation of the quality of observations.  Because cloud cover and other meteorological factors can impact image quality, observations may be negatively biased and underrepresent true NDVI if vegetation is obscured.  No weighting for the data were performed on the theory that the variation of these errors will be internalized by district and year fixed-effects.

\subsubsection{Rainfall Data}
To address the potential endogeneity of NDVI, we employ an instrumental variable strategy using rainfall as an exogenous driver of agricultural productivity. Rainfall data are sourced from the Climate Hazards Group InfraRed Precipitation with Station data (CHIRPS) dataset. CHIRPS provides quasi-global, high-resolution (0.05° x 0.05°) daily precipitation estimates from 1981 to the present by combining satellite imagery with in-situ station data, making it a reliable source for granular, historical weather patterns across Africa.

Evaluations of CHIRPS dataset against other sources, including in situ rainfall data collection, affirmed its accuracy and efficiency in estimating rainfall across diverse Rwandan terrains and thus may be used as a reliable source for rainfall data. However, the CHIRPS dataset was found to have slightly overestimated the amount of low-intensity rainfall (below 100mm) and slightly underestimated the amount of high-intensity rainfall (over 100mm), potentially creating a source of error (Kazora, et al., 2021).

\subsubsection{Education Data}
The primary response variables of interest are total school enrollment numbers for primary and secondary students by gender.  The following variables were taken from annual Rwanda Ministry of Education survey results and utilized in the final analysis (all the following are measured in each district and each year from 2010-2023): number of students (male, female, total) number of schools, number of classrooms, number of staff (male, female, total). All data are collected through self-reporting questionnaires and annual education censuses taken at the beginning of the academic year. Data quality assessments are performed in the latter half of the school year.

It is important to note that up until 2020, the Rwandan school year started in January and ended in November of the same year for primary and secondary schools. Beginning in 2020, the school calendar was adjusted to a September-June academic year for the primary and secondary levels. To maintain consistency in the analysis, the education data are associated with the year the school year began throughout the studied time period. This approach has the potential to cause inconsistencies in the results and be a source of error. Further, the staff data do not differentiate between qualifications at the district level. The observed values include both ‘qualified’ and ‘trained’ teachers. Additional notes and definitions related to data collection are included in the source reports.

\subsection{Descriptive Statistics}
Table 1 presents the summary statistics for the key variables used in the analysis. The data reveal several notable patterns. Primary school enrollment is substantial, with an average of over 42,000 male and 42,000 female students per district, reflecting Rwanda's success in achieving near-universal primary education. At the secondary level, enrollment is lower, and interestingly, female enrollment (mean of 11,278) is, on average, higher than male enrollment (mean of 9,678) across the districts in the sample. This reflects the positive impact of national gender equality initiatives in education.2 The standard deviations for all variables indicate considerable variation across districts and over time, which is essential for the fixed-effects identification strategy employed.

\begin{tabularx}{\linewidth}{l*{4}{Y}}
        \toprule
        \multicolumn{5}{l}{\textbf{Table A: Descriptive Statistics}} \\
        \midrule
               & Mean & Median & Standard Deviation & Observations \\[0pt]
               &      &        &    &     \\
    Primary Schools & 102.0095 & 98   & 31.4961 & 420      \\
            &      &        &     &    \\
    Primary Classrooms &  1184.8929 &  1104.5   & 354.1506 & 420      \\
        &      &        &     &    \\

    Primary Total Staff & 1621.9238 &  1509.5   & 476.3761 & 420      \\
            &      &        &    &     \\

    Primary Male Enrollment &  42491.5310 &  41560   & 9136.3484 & 420      \\
            &      &        &         \\
    Primary Female Enrollment &  42521.4857 &  41755   & 8970.0814 & 420      \\
        \midrule

    Secondary Schools & 54.9643 & 53   & 13.5323 & 420     \\
            &      &        &         \\
    Secondary Classrooms &  554.5024 &  563   & 138.1666   & 420   \\
        &      &        &         \\

    Secondary Total Staff &  909.0333 & 915  & 165.9329 & 390      \\
            &      &        &     &    \\

    Secondary Male Enrollment &  9677.6256 &  9478   & 1997.6637 & 390      \\
            &      &        &    &     \\
    Secondary Female Enrollment & 11278.3487  &  10976   & 2757.8415 & 390      \\

         \midrule

    Annual Rainfall (mm) & 1204.5608 & 1158.8750   & 332.3873 & 690      \\
            &      &        &     &    \\
    NDVI &  0.5688 &  0.5682   & 0.0343 & 690    
\end{tabularx}

\begin{center}
    \textit{Enrollment data are disaggregated by district, and year. Time periods for the collected observations are outlined in the Data section.}
\end{center}

It is both important and interesting to note that women have historically had more representation in education than men in non-tertiary education.  While this disparity has been diminished in primary enrollment, it has grown even more pronounced in secondary education. In 2023, every single one of the thirty districts had a female-to-male secondary enrollment ratio greater than one; however, only one of the thirty districts (Nyabihu) had a female-to-male \textit{primary} enrollment ratio greater than one in 2023 - down from twenty-six districts in 2010 - though all were above 0.9 and the median was 0.96. This trend reflects the impact of education initiatives which saw the net enrollment rate for girls in primary schools rise from 69.7\% in 1997 to 97.0\% in 2006. Boys primary enrollment has also now caught up (Ministry of Education, 2008).

The high primary enrollment rates at the start of – and throughout – the period of study helps control for biases caused by significant relative enrollment rate shocks.  Instead, changes in overall enrollment can thus be mostly explained by demographic distributions.  This creates a useful basis to investigate how exogenous agricultural shocks, and the resulting income and substitution effects, impact enrollment changes.

Figures \ref{fig:MeanNDVI} through \ref{fig:SecStaffRatio} in Appendix A show district-level changes in NDVI, total enrollment, gender enrollment ratios, total staff, and staff gender ratios over time.

\subsection{Identification Strategy}
The primary empirical challenge in estimating the causal effect of agricultural productivity on school enrollment is endogeneity. Unobserved, time-varying district-level factors—such as the quality of local governance, shifts in parental preferences for education, or the emergence of non-agricultural economic opportunities—may be correlated with both agricultural conditions (e.g., through investment in inputs) and educational outcomes, leading to biased estimates in an ordinary least squares (OLS) regression.

To overcome this challenge and isolate a causal relationship, this paper employs a two-stage least squares (2SLS) instrumental variable (IV) model, augmented with district and year fixed effects. This strategy uses plausibly exogenous, within-district variation in annual rainfall to instrument for our measure of agricultural productivity, NDVI. The core identifying assumption, or exclusion restriction, is that conditional on district and year fixed effects and time-varying school-level controls, annual rainfall affects school enrollment only through its impact on agricultural productivity and the associated household income shocks. While extreme rainfall events could theoretically affect enrollment through other channels, such as by damaging school infrastructure or roads, such widespread events are likely to be captured by the year fixed effects, which control for all nationwide shocks.

The econometric model is specified as follows:
\begin{itemize}
    \item[i)]  First Stage:
    \begin{equation}
        NDVI_{d,t,l} = \beta_0 + \beta_1\cdot \log{(Rainfall_{d,t})} + \vec{\beta_2} \cdot \chi_{d,t,l} + \varphi_t + \mu_d +\varepsilon_{d,t,l}
    \end{equation}
    \item[ii)] Second Stage:
    \begin{equation}
        \log{(Enrollment_{d,t,l})} = \beta_0 + \beta_1\cdot \widehat{NDVI}_{d,t,l} + \vec{\beta_2} \cdot \chi_{d,t,l} + \varphi_t + \mu_d +\varepsilon_{d,t,l}
    \end{equation}
    Where:
    \item $\log{(Enrollment_{d,t,l})}$ is the natural logarithm of the student population of interest (male or female) in district $d$, year $t$, and level $l$ (primary or secondary)
    \item $NDVI_{d,t}$ is the average Normalized Difference Vegetation Index for district $d$ in year $t$
    \item $\chi_{d,t,l}$ is a vector of time-varying school input controls, including the natural log of the number of schools, classrooms, and total staff in district $d$ for year $t$ and level $l$
    \item $\widehat{NDVI}_{d,t,l}$ is the predicted value of NDVI from the first-stage regression using the input controls from level $l$ (primary or secondary)
    \item $\mu_d$ is the district fixed effects, which absorb all time-invariant unobserved heterogeneity across districts, such as topography, baseline economic structure, and deep-seated cultural norms
    \item $\varphi_t$ is the year fixed effects, which control for any nationwide shocks or trends, such as national policy changes or macroeconomic fluctuations that affect all districts in a given year
    \item $\varepsilon_{d,t,l}$ is the error term
\end{itemize}

The validity of the 2SLS strategy hinges on the strength of the instrument. Table 2 presents the results of the first-stage regression of NDVI on rainfall, including the relevant school-level controls and fixed effects. Three different specifications for the rainfall term were used.

\begin{center}
\scalebox{0.7}{
\begin{tabularx}{0.7\linewidth}{l*{4}{Y}}
        \toprule
        \multicolumn{4}{l}{\textbf{Table B: Stage One Regression Results for Primary School}} \\
        \midrule
               & (1) & (2) & (3)\\[0pt]
               &      &        &         \\
    Log Rainfall Term Coefficient & $0.0573^{***}$ &  $0.0733^{***}$  &  $0.0438^{***}$     \\
    Standard Error & 0.0090 & 0.0092 & 0.0095    \\
    F-Statistic (Robust)   &  16.979    &  23.263      &   11.857      \\
    R-Squared  & 0.1496 & 0.1942 & 0.1094 \\
    Primary School Controls &  Y &  Y   & Y
\end{tabularx}
\begin{tabularx}{0.7\linewidth}{l*{4}{Y}}
        \toprule
        \multicolumn{4}{l}{\textbf{Table C: Stage One Regression Results for Secondary School}} \\
        \midrule
               & (1) & (2) & (3)\\[0pt]
               &      &        &         \\
    Log Rainfall Term Coefficient & $0.0370^{***}$ &  $0.0545^{***}$  &  $0.0198^{**}$     \\
    Standard Error & 0.0092 & 0.0094 & 0.0099    \\
    F-Statistic (Robust)   &  15.774    &  20.652      &   12.359      \\
    R-Squared  & 0.1496 & 0.1883 & 0.1219 \\
    Secondary School Controls &  Y &  Y   & Y
\end{tabularx}}\\
    \begin{tiny}
    \textit{Specification for Rainfall Term: (1) uses the amount rainfall in a calendar year; (2) uses rainfall amount from the beginning of the previous year's December to the beginning of the associated year's December; (3) uses rainfall amount from the beginning of the previous year's October to the beginning of the associated year's October.\\ $^{*}$ denotes significance at 10\% level; $^{**}$ denotes significance at 5\% level; $^{***}$ denotes significance at 1\% level}
    \end{tiny}
\end{center}

The results show a strong and statistically significant positive relationship between rainfall term (2) and NDVI. Crucially, the robust F-statistics for the excluded instrument are 23.261 for the primary school specification and 20.652 for the secondary school specification. As these values are well above the conventional rule-of-thumb threshold of 10 for a strong instrument, concerns about weak instrument bias are substantially mitigated, lending credibility to the causal interpretation of the 2SLS estimates. Rainfall term (2) is used for all subsequent analyses.

The validity of the exclusion restriction for the rainfall instrument is the primary methodological vulnerability of this analysis. Exploratory results included in Table 5 and Appendix B suggest the effect of rainfall does not encompass the effect of agriculture productivity, however the degrees to which this is true cannot be fundamentally verified. Certain endogenous effects, such as incidence of vector-borne or water-borne illnesses, localized infrastructure damage, and changes in time allocation related to water accessibility may create bias in the final results. Though some of the effects may be internalized by the year fixed effects, the author recognizes this may not by fully uncorrelated with the error terms in the preliminary regression.

\section{Results}
\subsection{Primary School Enrollment: A Dominant Substitution Effect}

The analysis of primary school enrollment reveals a countercyclical pattern: positive agricultural productivity shocks are associated with a decrease in enrollment, particularly in the years following the shock. This finding provides strong evidence that for basic education, where direct costs are low and enrollment is near-universal, household decisions are governed by a dominant substitution effect related to the opportunity cost of child labor.

\subsubsection{Main Results: A Countercyclicar Relationship}
Table 3 presents the core 2SLS regression results for primary school enrollment. The models estimate the impact of contemporaneous and lagged NDVI shocks on the natural logarithm of male, female, and total enrollment. The contemporaneous effect of an NDVI shock on enrollment is negative across all specifications but is not statistically distinguishable from zero. For total primary enrollment, the coefficient is -0.6584 with a p-value well above conventional significance levels. This initial lack of a significant response suggests that households possess short-term coping mechanisms that allow them to smooth both consumption and labor allocation in the immediate aftermath of a single-period shock. Families may draw down on savings, receive private transfers, or utilize informal credit to weather a single good or bad harvest without making significant adjustments to their children's schooling.

\subsubsection{Temporal Dynamics and Depleting Coping Mechanism}

The temporal dynamics of the household response are revealed more starkly when examining the impact of lagged productivity shocks. The effect of NDVI on primary enrollment becomes substantially larger in magnitude and gains high statistical significance in the years following the initial shock. As shown in Table 3, the coefficient on the one-year lagged NDVI for total enrollment is -0.8930 and is significant at the 5\% level. The effect intensifies further with a two-year lag, where the coefficient is -1.7316 ($p<0.01$), indicating that a 0.0343 unit increase in NDVI, which portends a year's crop production one standard deviation above the mean, two years prior is associated with a 5.8\% decrease in current primary school enrollment. The effect remains large and highly significant for the three-year lag (-1.2010, $p<0.01$).

%Above I used the equation: $e^{-1.7316*0.0343} - 1$ to obtain the interpretation. I used 0.0343 because it is the standard deviation of the entire NDVI dataset. Gemini, is this correct? What else should I do instead?

This robust temporal pattern is a central finding of this study. It suggests a process whereby household resilience is eroded by the persistence of shocks. While a family may be able to absorb a single adverse agricultural shock (low NDVI) using buffer stocks or other short-term strategies, a sustained period of low productivity compels more drastic adjustments. As savings are depleted and credit constraints bind more tightly, the economic calculus shifts. A prolonged agricultural downturn drastically reduces the marginal product of child labor on the farm, causing the opportunity cost of education to collapse. In this environment, sending a child to school becomes the more economically rational choice for the household, leading to an increase in enrollment. Conversely, a sustained period of high productivity (high NDVI) persistently raises the opportunity cost of schooling, leading to a sustained decrease in enrollment. This lagged response provides strong evidence that human capital investment decisions are part of a multi-period optimization problem where households adapt to persistent changes in their economic environment.

\subsubsection{Gendered Responses: A Challenge to the ``Girls as a Buffer'' Hypothesis}
A particularly noteworthy finding emerges when disaggregating the lagged effects by gender. Contrary to much of the literature, which often identifies girls as the primary margin of adjustment for household labor needs, the results for Rwanda indicate that boys' primary school enrollment is more sensitive to agricultural productivity shocks.

As detailed in Table 3, at a one-year lag, the coefficient for male enrollment is -1.0403 ($p<0.05$), while the coefficient for female enrollment (-0.7506) is smaller and not statistically significant. This gap widens dramatically at the two-year lag, with a coefficient of -2.3095 ($p<0.01$) for males compared to -1.1564 ($p<0.01$) for females. The effect for males is double that for females. This pattern, where the male coefficient is consistently larger in magnitude and often more statistically significant, persists across all lagged specifications.

This result challenges the universality of the ``girls as a buffer'' hypothesis and underscores the critical importance of context-specific gender roles in production. The greater elasticity of boys' enrollment to NDVI suggests that their labor is more substitutable with schooling in direct response to agricultural conditions. This implies that in the Rwandan context, primary-school-aged boys may be more intensively involved in agricultural tasks whose marginal productivity is highly sensitive to crop yields—such as planting, weeding, harvesting, or managing livestock (Jensen, 2010). Girls of a similar age, conversely, may be allocated to domestic production tasks (e.g., childcare, fetching water, food preparation) that are essential for household maintenance but whose value is less directly tied to the success of the harvest (Sawadogo and Mabugu, 2025).

Therefore, during a positive agricultural shock (a high NDVI year), the opportunity cost of a boy's time rises sharply, creating a strong incentive to pull him from school to contribute to farm production. During a negative shock, this high opportunity cost dissipates, making his return to school a more efficient allocation of household resources. This presents a nuanced picture that contrasts sharply with findings from neighboring Uganda, demonstrating significant heterogeneity in intra-household, gendered responses to income shocks even within the same broader geographic and economic region.

\begin{center}
\begin{tabularx}{\linewidth}{l*{4}{Y}}
        \toprule
        \multicolumn{4}{l}{\textbf{Table 3: Stage Two Regression Results for Primary Enrollment}} \\
        \midrule
               & Log Primary Male Enrollment & Log Primary Female Enrollment & Log Primary Total Enrollment \\[0pt]
               &      &        &         \\
        $\widehat{NDVI}$ & $-0.5923$	& $-0.7259$	& $-0.6584$ \\
        & (0.4301)	& (0.4638)	& (0.4407) \\
                    &      &        &         \\
        $\widehat{NDVI}$ Lag One Year & $-1.0403^{**}$	& $-0.7506^{*}$	& $-0.8930^{**}$ \\
        & (0.4109)	& (0.4445)	& (0.4214) \\
                    &      &        &         \\
        $\widehat{NDVI}$ Lag Two Years & $-2.3095^{***}$	& $-1.1564^{***}$	& $-1.7316^{***}$ \\
        & (0.3619)	& (0.4048)	& (0.3788) \\
                    &      &        &         \\
        $\widehat{NDVI}$ Lag Three Years & $-1.4876^{***}$	& $-0.9226^{**}$	& $-1.2010^{***}$ \\
        & (0.3890)	& (0.4102)	& (0.3933) \\
                    &      &        &       \\
\end{tabularx}
\begin{tiny}
\textit{Notes: Each column represents a separate 2SLS fixed-effects regression. All regressions include district and year fixed effects as well as Primary education control variables\\ Standard errors are in parentheses\\ $^{*}$ denotes significance at 10\% level; $^{**}$ denotes significance at 5\% level; $^{***}$ denotes significance at 1\% level}
    \end{tiny}
\end{center}

\subsection{Secondary School Enrollment: Binding Constraints and the Income Effects}

The analysis of secondary school enrollment reveals a starkly different narrative, one characterized by a complete reversal of the relationship observed at the primary level. Here, positive agricultural shocks are strongly and positively associated with increased enrollment. This finding indicates that the income effect, driven by the high costs of post-primary education and the presence of binding credit constraints, is the dominant mechanism governing household investment decisions at this stage.

\subsubsection{Main Results: The Procyclical Reversal}

Table 4 presents the 2SLS regression results for secondary enrollment. The models show that contemporaneous NDVI shocks have a large, positive, and highly statistically significant effect on enrollment. The coefficient for total secondary enrollment is 5.1206 ($p<0.01$), a value that is not only opposite in sign but also an order of magnitude larger than the primary school coefficients. This result implies that a 0.0343 unit increase in NDVI, which portends a year's crop production one standard deviation above the mean, is associated with a remarkable 19.2\% increase in secondary enrollment in the same year. The effect remains strongly positive and significant for the one-year lag (6.4358, $p<0.01$) and the two-year lag (1.6461, $p<0.01$).
%Above I used the equation: $e^{5.1206*0.0343} - 1$ to obtain the interpretation. I used 0.0343 because it is the standard deviation of the entire NDVI dataset. Gemini, is this correct? What else should I do instead?

This dramatic reversal from the primary school results points unequivocally to a dominant income effect. Unlike primary education, which has achieved near-universal enrollment in Rwanda and involves low direct costs for households, secondary education represents a substantial financial investment. The costs include not only higher potential fees, uniforms, and materials but also a significantly higher opportunity cost, as the forgone earnings of an adolescent are much greater than those of a young child.

In the presence of imperfect or nonexistent credit markets, many agrarian households are constrained in their ability to finance such large, lumpy investments. Consequently, secondary education functions as a normal good: households can only afford to purchase it when their income rises (Mimoun, 2008). A positive agricultural shock provides the necessary liquidity to overcome these financial barriers, relaxing the credit constraint and leading to a surge in enrollment. This finding is consistent with a broad literature showing that financial barriers are a primary determinant of secondary school enrollment in developing countries (Lochner and Monge-Naranjo, 2012).

\subsubsection{Evolving Gender Asymmetries and Long-Term Strategy}

While the initial response to positive shocks is strongly positive for both genders, the temporal dynamics reveal complex and evolving gender asymmetries that suggest a multi-period decision-making process that goes beyond a simple, static trade-off. The contemporaneous effect is large and positive for both males (4.8338, $p<0.01$) and females (5.3341, $p<0.01$). However, the pattern diverges over time.

For males, the positive effect remains highly significant through the two-year lag before becoming small and statistically insignificant at the three-year lag (-0.5073, $p>0.1$). The trajectory for females is even more dramatic. The positive effect weakens at the two-year lag, and by the three-year lag, the coefficient becomes large, negative, and highly statistically significant (-1.7394, $p<0.01$). This indicates that a sustained period of agricultural prosperity ultimately leads to a significant decrease in female secondary school enrollment three years later, an effect that is not observed for their male counterparts.

This dynamic pattern suggests that persistent prosperity may alter the economic landscape and household strategies in a gender-differentiated manner. The initial positive effect confirms the dominance of the income effect for both genders, as households use the windfall to overcome immediate financial hurdles. The subsequent divergence, particularly the strong negative turn for females, implies that a sustained period of high income triggers a re-evaluation of the long-term relative returns to different life paths for daughters. Several mechanisms could explain this phenomenon.

One researched explanation suggests that sustained household wealth may improve a daughter's marriage market prospects, leading the family to prioritize an early, advantageous marriage over further educational attainment, a pattern observed in certain low-income settings (Zhou, 2024). In the context of Rwanda, however, the minimum age of marriage is 21 and thus matrimony is unlikely to contribute to girls' secondary school enrollment divergence.

Another explanation is that a booming local economy, driven by agricultural prosperity, may create new employment opportunities in non-agricultural sectors (e.g., local markets, service-related work), making a teenage daughter's immediate labor more valuable to the household than the uncertain future returns from completed schooling (Elmallakh and Wodon, 2022).

The fact that the effect for males does not turn significantly negative suggests that the perceived long-term economic returns to male secondary education may be sufficiently high to counteract the rising opportunity cost of their time. This evolving, gendered response to the \textit{persistence} of shocks is a significant finding, demonstrating that household investment strategies are not static but adapt dynamically to changing long-term economic outlooks.

\begin{center}
\begin{tabularx}{\linewidth}{l*{4}{Y}}
        \toprule
        \multicolumn{4}{l}{\textbf{Table 4: Stage Two Regression Results for Secondary Enrollment}} \\
        \midrule
               & Log Secondary Male Enrollment & Log Secondary Female Enrollment & Log Secondary Total Enrollment \\[0pt]
               &      &        &         \\
        $\widehat{NDVI}$ & $4.8338^{***}$	& $5.3341^{***}$	& $5.1206^{***}$ \\
        & (0.8247)	& (0.8633)	& (0.8016) \\
                    &      &        &         \\
        $\widehat{NDVI}$ Lag One Year & $6.5754^{***}$	& $6.3122^{***}$	& $6.4358^{***}$ \\
        & (0.6084)	& (0.6785)	& (0.6091) \\
                    &      &        &         \\
        $\widehat{NDVI}$ Lag Two Years & $1.7495^{***}$	& $1.5858^{**}$	& $1.6461^{***}$ \\
        & (0.5853)	& (0.6543)	& (0.5964) \\
                    &      &        &         \\
        $\widehat{NDVI}$ Lag Three Years & $-0.5073$	& $-1.7394^{***}$	& $-1.1675^{**}$ \\
        & (0.5681)	& (0.6517)	& (0.5891) \\
                    &      &        &
\end{tabularx}
\begin{tiny}
    \textit{Notes: All regressions include district and year fixed effects as well as secondary education control variables\\ Standard errors are in parentheses\\ $^{*}$ denotes significance at 10\% level; $^{**}$ denotes significance at 5\% level; $^{***}$ denotes significance at 1\% level}
\end{tiny}
\end{center}

\section{Discussion and Synthesis}

The empirical results present a nuanced and multi-faceted account of how agrarian households in Rwanda navigate the trade-offs between agricultural production and human capital investment. The findings, when synthesized, paint a cohesive picture of household decision-making as a multi-layered and state-dependent process, where the logic governing investment shifts fundamentally depending on the costs, benefits, and constraints associated with each level of schooling.

\subsection{A Unified, State-Dependent Model of Household Investment}

The core synthesis of this paper's findings lies in the direct comparison of the results for primary and secondary schooling. The evidence strongly supports a bifurcated model of household decision-making. For primary education, where Rwanda has achieved near-universal enrollment and direct costs are low due to fee-free policies, the binding constraint on a household's decision is the opportunity cost of the child's time. In this regime, the substitution effect dominates. Agricultural downturns lower the marginal product of child labor on the farm, reducing this opportunity cost and making schooling a relatively more attractive use of a child's time. This leads to the observed countercyclical enrollment pattern, where enrollment rises in response to persistent negative productivity shocks.

For secondary education, which has a lower transition rate and involves significantly higher direct and indirect costs, the binding constraint shifts from opportunity cost to financial capacity. Here, the income effect dominates. Households, facing liquidity and credit constraints, can only make the substantial investment required for secondary schooling during periods of increased income. This leads to a strongly procyclical enrollment pattern, where good harvests translate directly into higher enrollment. This dichotomy provides robust empirical support for the theoretical framework articulated by Ferreira \& Schady (2008), which posits that the cyclicality of educational investment depends on the relative strength of income and substitution effects, which can vary across contexts and levels of development.

Furthermore, the inclusion of multiple lagged variables allows for an analysis of how households adapt over a multi-year horizon. At the primary level, the strengthening of the substitution effect over time suggests a process of depleting buffer stocks. Households appear to resist adjusting school enrollment after a single bad harvest but are forced to do so as the shock's effects linger, revealing the limits of their consumption-smoothing capacity. At the secondary level, the dynamic is more complex, suggesting a re-evaluation of long-term strategies. A short-term positive shock is treated as a windfall to be invested in a high-return asset (secondary education). However, a persistent positive shock appears to change households' fundamental expectations about the economic environment. This leads to a reassessment of the relative returns of different life paths (e.g., continued schooling versus household production for girls, or schooling versus immediate employment for boys), resulting in the observed reversal of the enrollment effect for females in the third year after the initial shock.

\subsection{Interpreting Endogenous Policy Responses: The Case of School Inputs}

A consistent and initially puzzling finding across several of the underlying OLS models is the negative and statistically significant coefficient on the logarithm of total school staff. Table 6 in Appendix B presents the full regression results using NDVI and the control variables. A literal causal interpretation—that hiring more staff causes enrollment to decline—is economically implausible. Instead, this result is best understood as evidence of endogeneity arising from non-random program placement, a common challenge in estimating education production functions. The negative coefficient likely reflects a policy of compensatory resource allocation, whereby educational authorities direct additional resources, including teachers and support staff, to districts that are systematically underperforming.

Districts with unobserved, time-varying characteristics that lead to poor educational outcomes—such as high rates of poverty, low historical achievement, or inadequate infrastructure—are precisely the ones targeted for intervention. The fixed-effects specification controls for time-invariant district characteristics, but it cannot account for the fact that the allocation of new staff over time may be a direct response to negative trends in enrollment or performance. Therefore, the regression captures the correlation between being a struggling district (which receives more staff) and having lower enrollment, rather than the causal effect of staff on enrollment. This interpretation suggests that the Rwandan education system actively attempts to address regional disparities, though the effectiveness of this resource allocation is beyond the scope of this analysis.

\subsection{Sources of Bias and Areas of Future Research}

The data quality and aggregation bias may introduce sources of error. Because cloud cover and other meteorological factors can negatively bias observations in the MODIS dataset, the measured NDVI may underrepresent the true vegetation health and yield. Though district and year fixed effects may internalize part of the resulting error, this assumption requires further analysis. Similarly, the aggregation of the NDVI mean at the district level may obfuscate variance in Rwanda's diverse micro-climate setting. Using the district mean may smooth the variance and obscure intra-district household-level heterogeneity in agriculture production. Consequently, the estimated coefficients are likely conservative and understate the true behavioral response to income shocks at the household level.

Future research should prioritize household-level data and including specific health controls. Moving to household-level panel data is crucial for confirming these aggregate findings and precisely modeling the underlying mechanisms, such as intra-household bargaining dynamics, the differential allocation of time to specific agricultural and domestic tasks, and the exact nature of credit constraints faced by agrarian families. To robustly satisfy the exclusion restriction, future econometric analysis must incorporate health or disease outcomes correlated with rainfall variability, directly controlling for the non-agricultural channel of effect. Though related datasets are publicly available, each iteration of the Demographic and Health Survey as well as the Integrated Household Living Survey occurs quinquennially and triennially, respectively, and has not accumulated enough times to provide significant insight.

\subsection{Consolidated Results}

Table 5 provides a consolidated summary of the preliminary regression of NDVI, reduced form regression using rainfall, and estimated 2SLS effects of NDVI on log enrollment, facilitating a direct comparison of the key findings and providing a robustness check. The table starkly illustrates the paper's central narrative: the consistently negative coefficients for primary enrollment stand in sharp contrast to the initially positive coefficients for secondary enrollment. It also highlights the key gender dynamics: the larger magnitude of the negative effect for boys at the primary level and the dramatic negative turn for girls at the secondary level after a three-year lag.

\begin{center}
\scalebox{0.7}{
    \begin{tabularx}{\linewidth}{l*{5}{Y}}
        \toprule
        \multicolumn{5}{l}{\textbf{Table 5: Combined Results}} \\
        \midrule
        & \textbf{Primary}  & & \textbf{Secondary}  \\
        &      &        &    &     \\

        & Male & Female & Male  &  Female \\[0pt]
        \midrule
               &      &        &         \\
        NDVI &  $-0.9513^{***}$ &  $-0.8443^{***}$	 & 0.3607	& $0.4501^{*}$\\
        & (0.1554) & (0.1703) & (0.2523)	& (0.2651) \\
                    &      &        &         \\

        Log Annual Rainfall & $-0.0434$	& $-0.0532$	& $0.2634^{***}$	& $0.2907^{***}$ \\
        & (0.0315)	& (0.0340)	& (0.0449)	& (0.0470) \\
                    &      &        &         \\
        $\widehat{NDVI}$ & $-0.5923$	& $-0.7259$	& $4.8338^{***}$	& $5.3341^{***}$ \\
        & (0.4301)	& (0.4638)	& (0.8247)	& (0.8633) \\
        \midrule
        NDVI Lag One Year &  $-0.4183^{**}$ &  $-0.347^{**}$   & $0.4637^{*}$	& 0.3056      \\
        & (0.1660) & (0.1797) & (0.2553)	& (0.2692)   \\
                    &      &        &         \\
        Log Annual Rainfall Lag One Year & $-0.0587^{*}$	& $-0.0484$	& $0.4410^{***}$	& $0.4570^{***}$ \\
        & (0.0309)	& (0.0334)	& (0.0387)	& (0.0422) \\
                    &      &        &         \\
        $\widehat{NDVI}$ Lag One Year & $-1.0403^{**}$	& $-0.7506^{*}$	& $6.5754^{***}$	& $6.3122^{***}$ \\
        & (0.4109)	& (0.4445)	& (0.6084)	& (0.6785) \\
        \midrule
        NDVI Lag Two Years &  $-0.4331^{***}$ &  -0.2740   & $-0.4604^{**}$	& $-0.6255^{**}$      \\
        & (0.1645) & (0.1785) & (0.2429)	& (0.2547)    \\
        &      &        &         \\
        Log Annual Rainfall Lag Two Years & $-0.1611^{***}$	& $-0.0747^{**}$	& $0.1079^{**}$	& $0.1413^{***}$ \\
        & (0.0295)	& (0.0327)	& (0.0428)	& (0.0474) \\
                    &      &        &         \\
        $\widehat{NDVI}$ Lag Two Years & $-2.3095^{***}$	& $-1.1564^{***}$	& $1.7495^{***}$	& $1.5858^{**}$ \\
        & (0.3619)	& (0.4048)	& (0.5853)	& (0.6543) \\
        \midrule
        NDVI Lag Three Years &  $-0.4411^{***}$ &  $-0.5204^{***}$   & $-0.4065^{*}$	& $-0.8267^{***}$      \\
        & (0.1645) & (0.1785) & (0.2399)	& (0.2495)    \\
        &      &        &         \\
        Log Annual Rainfall Lag Three Years & $-0.0736^{**}$	& $-0.0405$	& $0.0484$	& $0.0433$ \\
        & (0.0302)	& (0.0315)	& (0.0471)	& (0.0547) \\
        &      &        &       \\
        $\widehat{NDVI}$ Lag Three Years & $-1.4876^{***}$	& $-0.9226^{**}$	& $-0.5073$	& $-1.7394^{***}$ \\
        & (0.3890)	& (0.4102)	& (0.5681)	& (0.6517) \\
        \bottomrule
        &      &        &       
\end{tabularx}}\\

\begin{tiny}
\textit{Notes: Each column represents a separate 2SLS fixed-effects regression. All regressions include district and year fixed effects as well as level-specific education control variables\\ Standard errors are in parentheses\\ $^{*}$ denotes significance at 10\% level; $^{**}$ denotes significance at 5\% level; $^{***}$ denotes significance at 1\% level}
\end{tiny}
\end{center}

\section{Conclusion}

This paper has investigated how households in Rwanda adjust their educational investments in children in response to exogenous agricultural productivity shocks. Using an instrumental variable strategy with district and year fixed effects, the analysis provides causal evidence on the mechanisms governing enrollment decisions for primary and secondary education. The findings reveal a bifurcated logic: primary school enrollment is countercyclical, driven by a dominant substitution effect related to the opportunity cost of child labor, while secondary school enrollment is strongly procyclical, governed by a dominant income effect in the face of binding financial constraints. The study also uncovers novel, context-specific gender dynamics, showing that boys' primary enrollment is more sensitive to agricultural shocks, and that girls' secondary enrollment declines in response to sustained periods of prosperity.

The findings yield several actionable policy implications for Rwanda and other developing countries with similar economic structures.
\begin{itemize}
    \item[i.] Supporting Secondary Enrollment: Given that secondary enrollment is highly sensitive to positive income shocks, policies should focus on alleviating the binding financial constraints that prevent households from investing. This could include targeted scholarships, conditional cash transfers that are triggered during agriculturally prosperous years, or expanded access to educational credit and savings programs. Such interventions would allow households to translate temporary income gains into long-term human capital investments, breaking the link between transitory poverty and educational attainment.

    \item[ii.] Stabilizing Primary Enrollment: The countercyclical nature of primary enrollment suggests that households pull children (especially boys) out of school when the opportunity cost of their labor is high. To mitigate this, policies should aim to smooth agricultural income and delink it from child labor allocation. This reinforces the case for agricultural insurance products, price stabilization mechanisms for key crops, and the promotion of off-farm income-generating activities that can provide a more stable income source, reducing the need to rely on child labor during peak agricultural seasons.

    \item[iii.] Gender-Aware Policy Design: Policies must be designed with an awareness of the specific, context-dependent gender dynamics at play. In this setting, interventions aimed at keeping primary-aged boys in school during good harvests may be particularly effective. At the secondary level, understanding the factors that drive girls to leave school after periods of sustained prosperity-such as alternative production opportunities—is critical for designing effective retention programs, such as mentorship initiatives or programs that clearly demonstrate the high long-term economic returns to female secondary education.
\end{itemize}

\newpage
\begin{appendices}
\section{Descriptive Figures}

\begin{figure}[h!]
    \centering
        \includegraphics[width=1\linewidth]{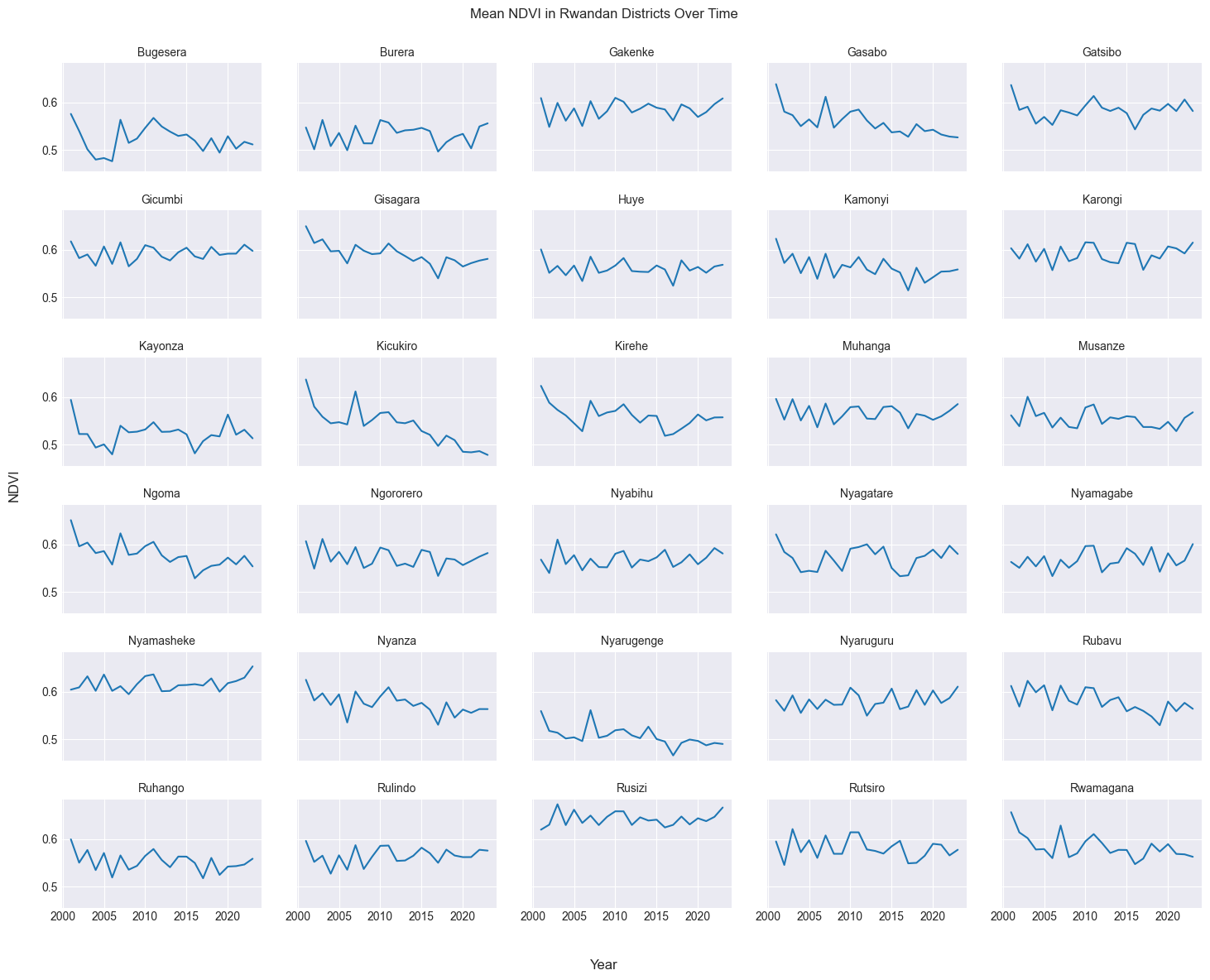}
        \captionof{figure}{Mean NDVI by District Over Time (2001 - 2023)}
        \label{fig:MeanNDVI}
\end{figure}

\begin{figure}[h!]
    \centering
   \begin{subfigure}{.5\textwidth}
   \centering
        \includegraphics[width=1\linewidth]{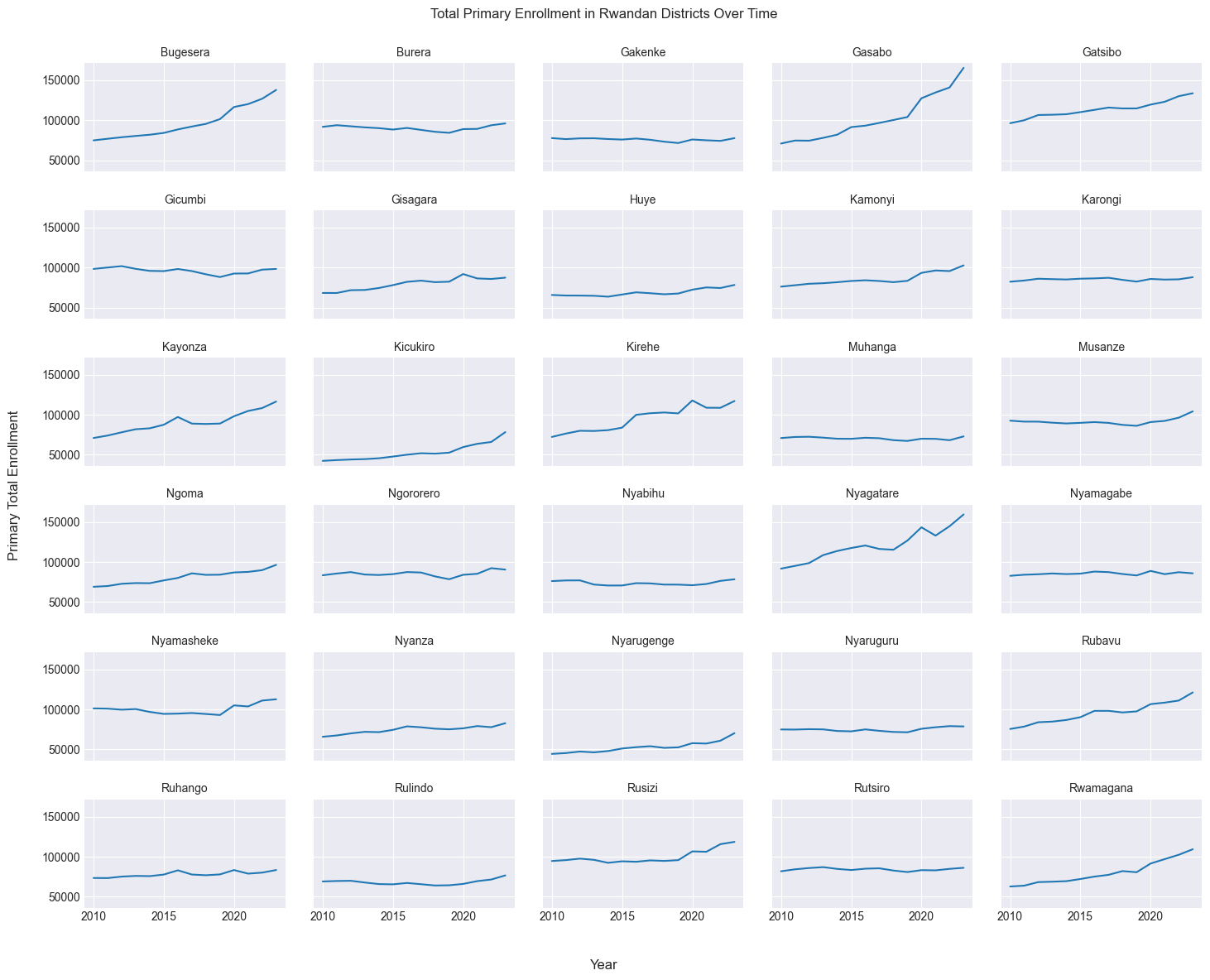}
        \caption{Primary Enrollment (2010 - 2023)}
        \label{fig:PrimEnroll}
    \end{subfigure}%
    \begin{subfigure}{.5\textwidth}
    \centering
        \includegraphics[width=1\linewidth]{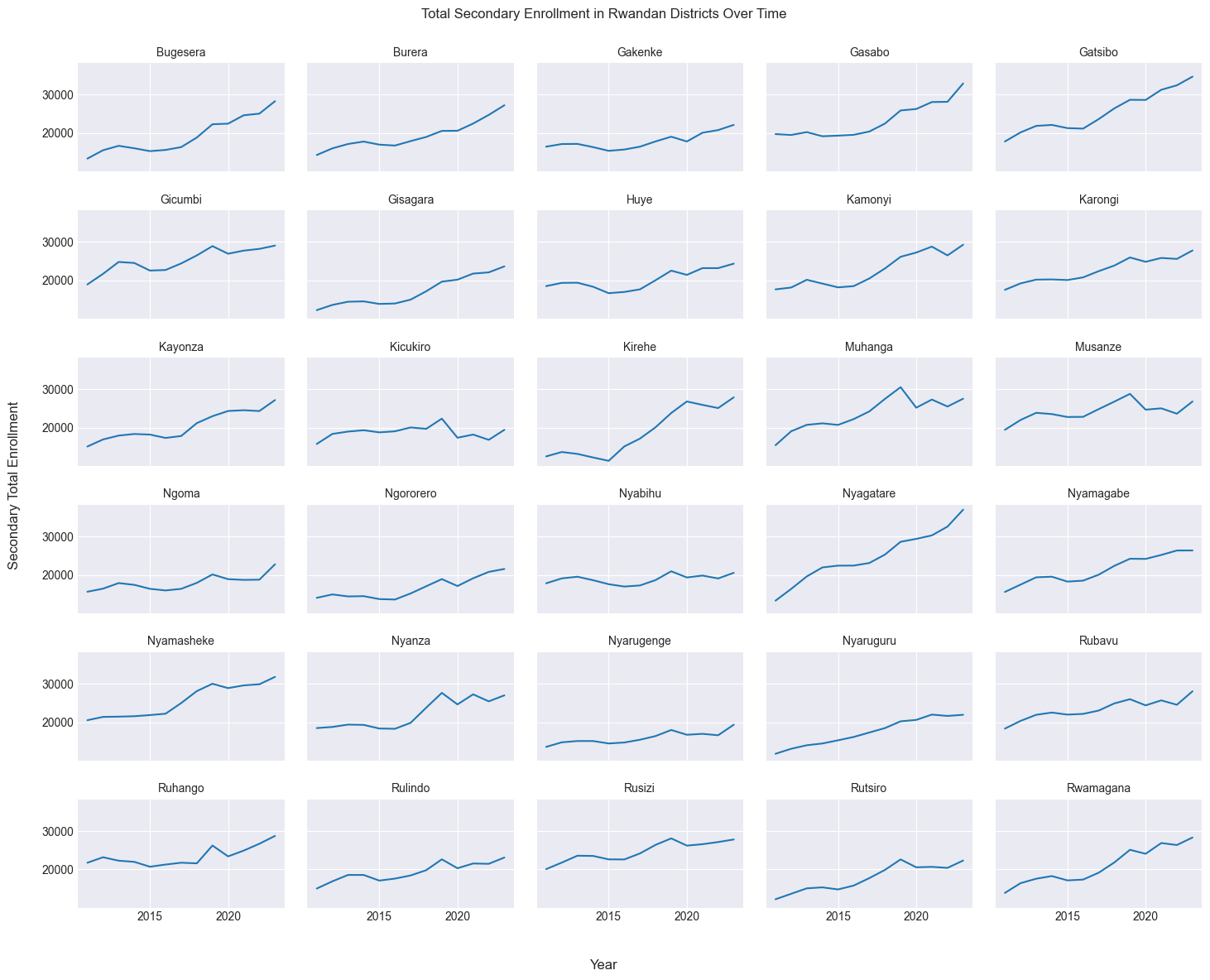}
        \caption{Secondary Enrollment (2011 - 2023)}
        \label{fig:SecEnroll}
    \end{subfigure}
    \caption{Total Primary and Secondary Enrollment by District Over Time}
\end{figure}

\begin{figure}[h!]
    \centering
   \begin{subfigure}{.5\textwidth}
   \centering
    \includegraphics[width=1\linewidth]{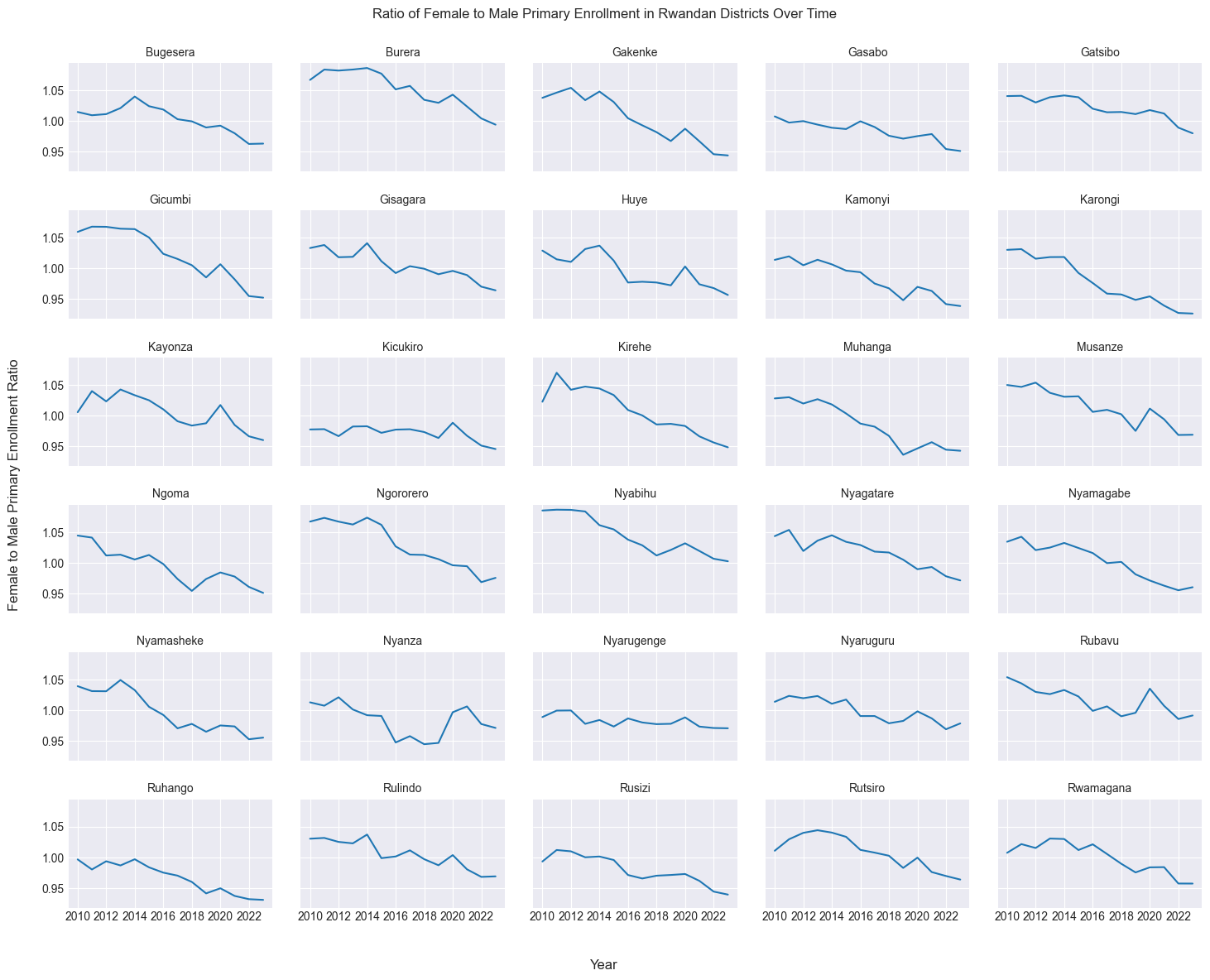}
    \caption{Primary Enrollment Gender Ratio (2010 - 2023)}
    \label{fig:PrimEnrollRatio}
    \end{subfigure}%
    \begin{subfigure}{.5\textwidth}
    \centering
    \includegraphics[width=1\linewidth]{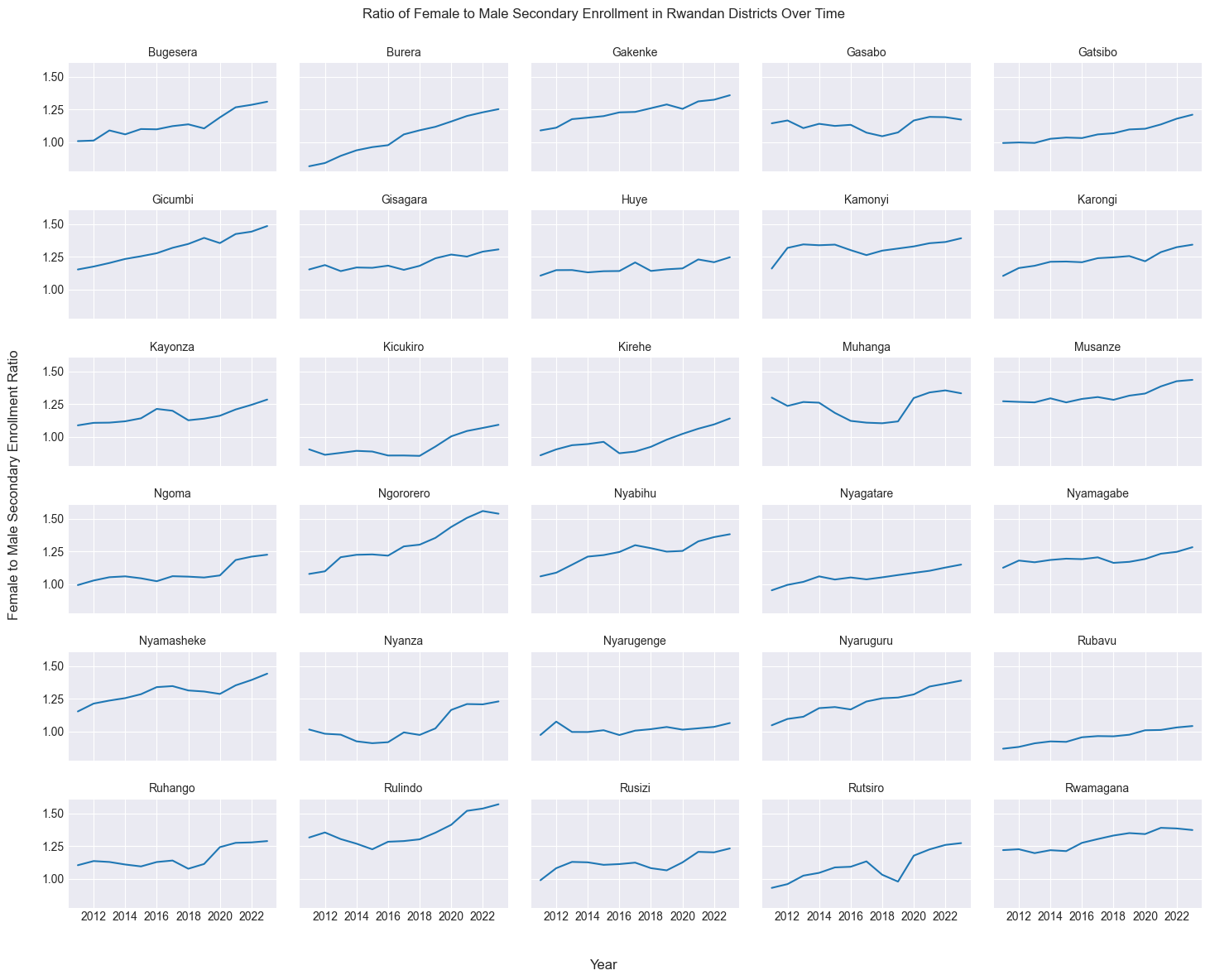}
    \caption{Secondary Enrollment Gender Ratio (2011 - 2023)}
    \label{fig:SecEnrollRatio}
    \end{subfigure}
    \caption{Female-to-Male Primary and Secondary Enrollment Ratio by District Over Time}
\end{figure}

\begin{figure}[h!]
    \centering
   \begin{subfigure}{.5\textwidth}
   \centering
     \includegraphics[width=1\linewidth]{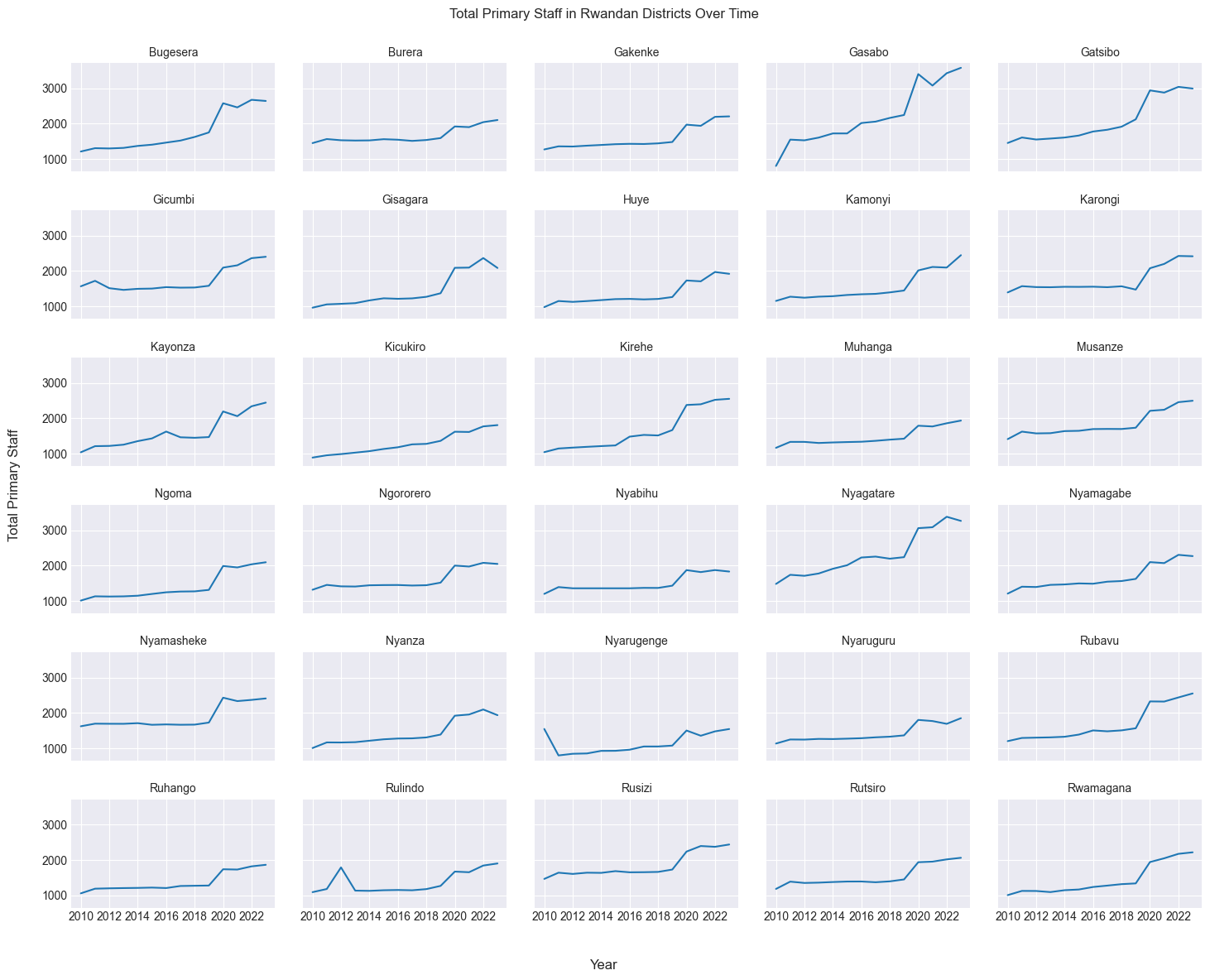}
    \caption{Total Primary Staff (2010 - 2023)}
    \label{fig:PrimStaff}
    \end{subfigure}%
    \begin{subfigure}{.5\textwidth}
    \centering
    \includegraphics[width=1\linewidth]{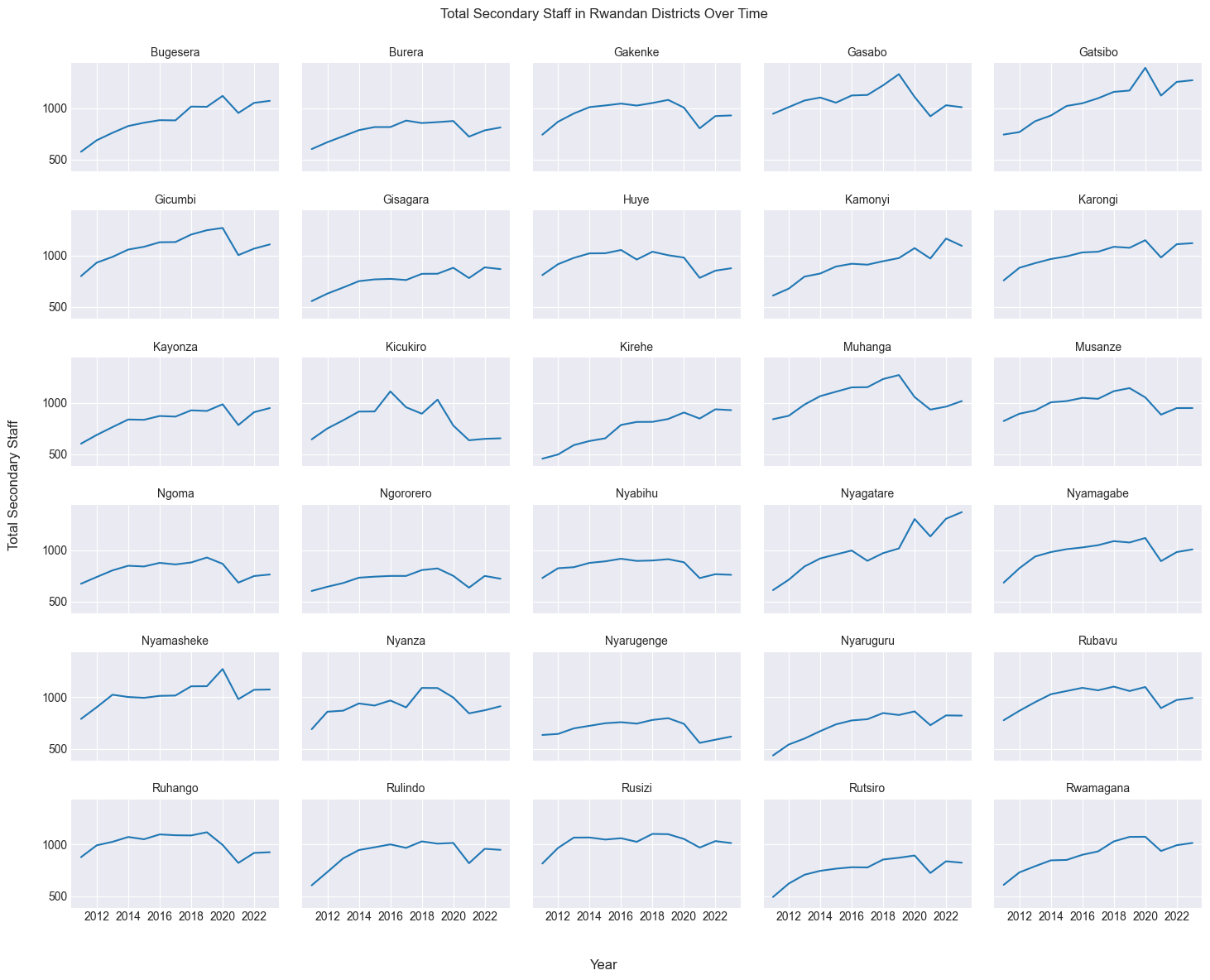}
    \caption{Total Secondary Staff (2011 - 2023)}
    \label{fig:SecStaff}
    \end{subfigure}
    \caption{Total Primary and Secondary Staff by District Over Time}
\end{figure}

\begin{figure}[h!]
    \centering
   \begin{subfigure}{.5\textwidth}
   \centering
     \includegraphics[width=1\linewidth]{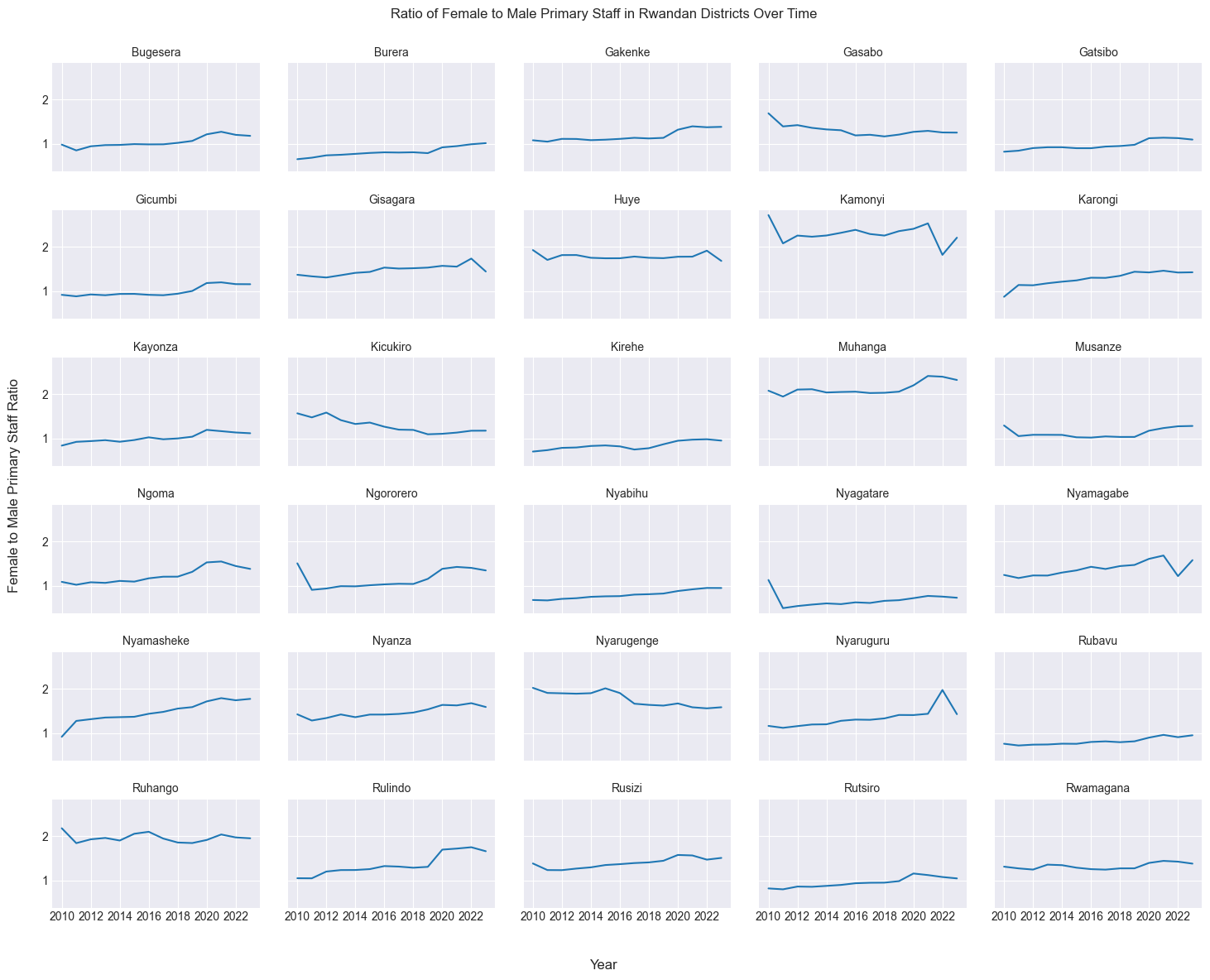}
    \caption{Primary Staff Gender Ratio (2010 - 2023)}
    \label{fig:PrimStaffRatio}
    \end{subfigure}%
    \begin{subfigure}{.5\textwidth}
    \centering
    \includegraphics[width=1\linewidth]{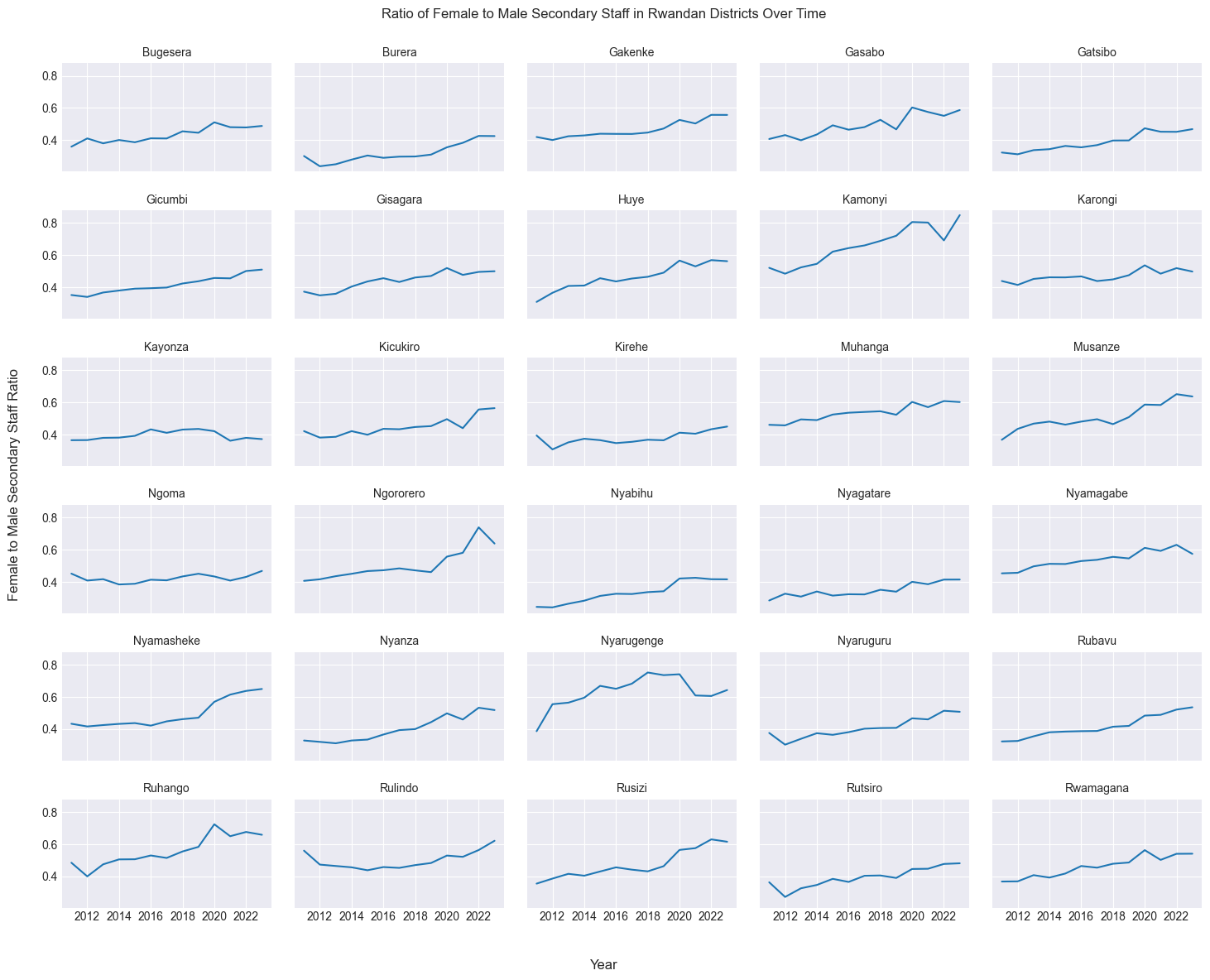}
    \caption{Secondary Staff Gender Ratio (2011 - 2023)}
    \label{fig:SecStaffRatio}
    \end{subfigure}
    \caption{Female to Male Primary and Secondary Staff Ratio by District Over Time}
\end{figure}
\clearpage
\section{NDVI Relationship Figures}
\begin{figure}[h!]
    \centering
        \includegraphics[width=0.9\linewidth]{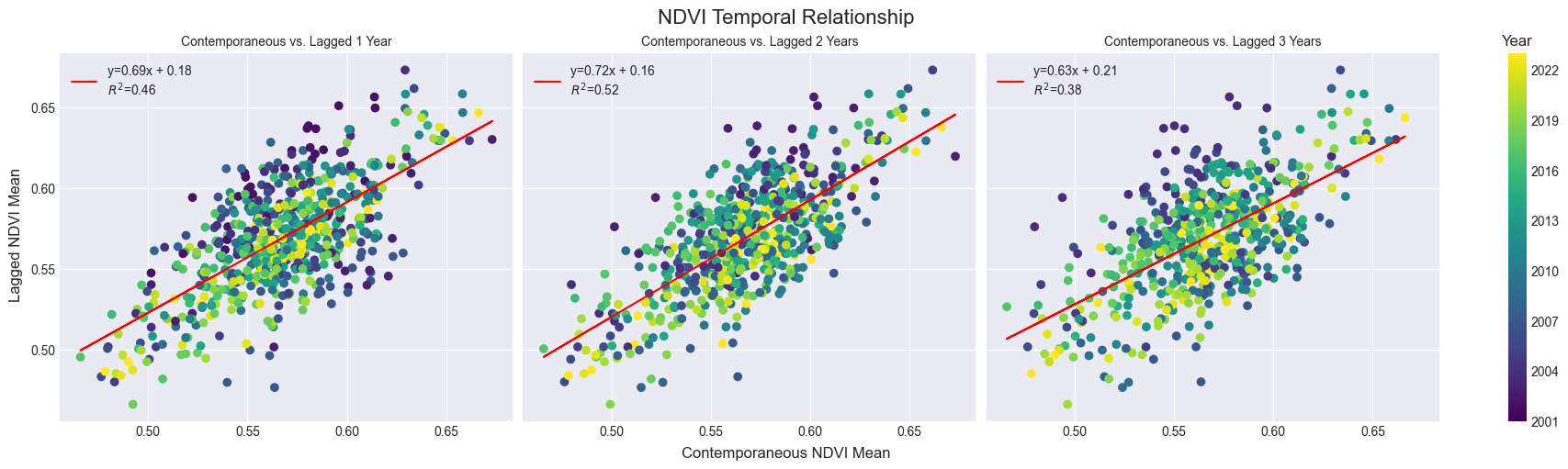}
        \captionof{figure}{True NDVI Temporal Relationship (2001 - 2023)}
        \label{fig:NDVITemp}
\end{figure}

\begin{figure}[h!]
    \centering
        \includegraphics[width=0.9\linewidth]{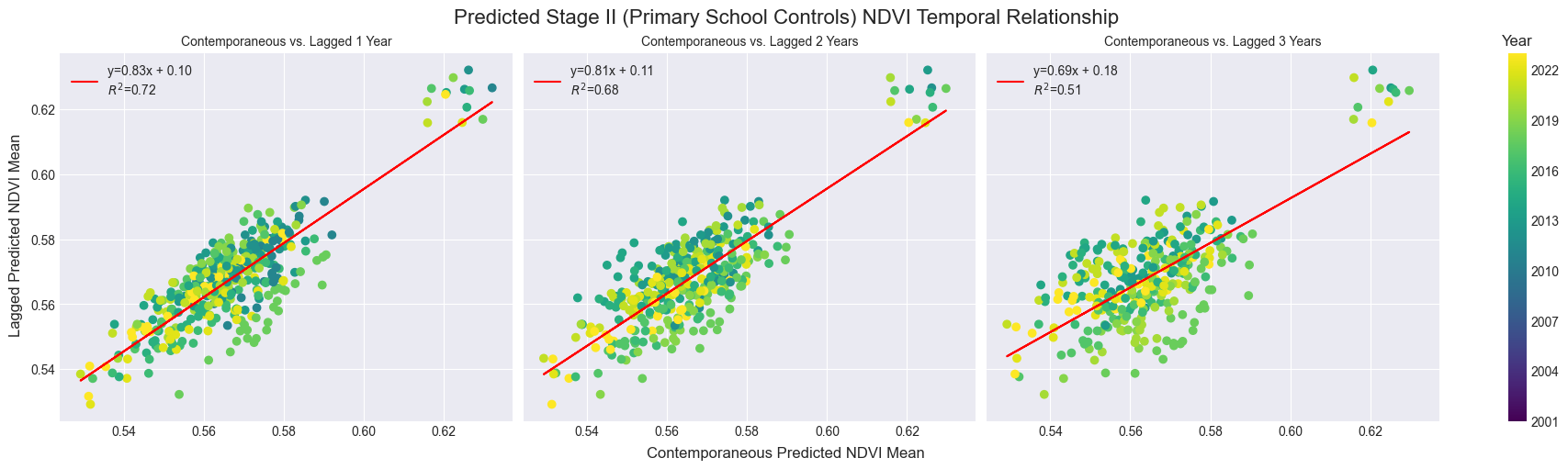}
        \captionof{figure}{Predicted Stage II NDVI (Primary School Controls) Temporal Relationship (2001 - 2023)}
        \label{fig:PredPrimNDVITemp}
\end{figure}

\begin{figure}[h!]
    \centering
        \includegraphics[width=0.9\linewidth]{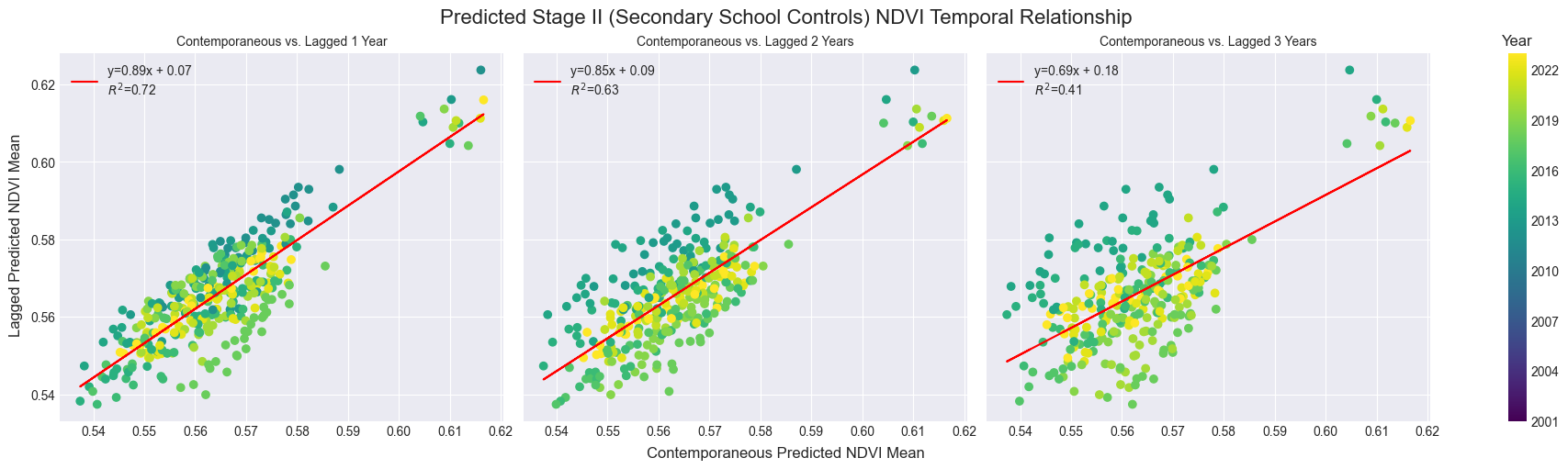}
        \captionof{figure}{Predicted Stage II NDVI (Secondary School Controls) Temporal Relationship (2001 - 2023)}
        \label{fig:PredSecNDVITemp}
\end{figure}

\begin{figure}[h!]
    \centering
        \includegraphics[width=0.7\linewidth]{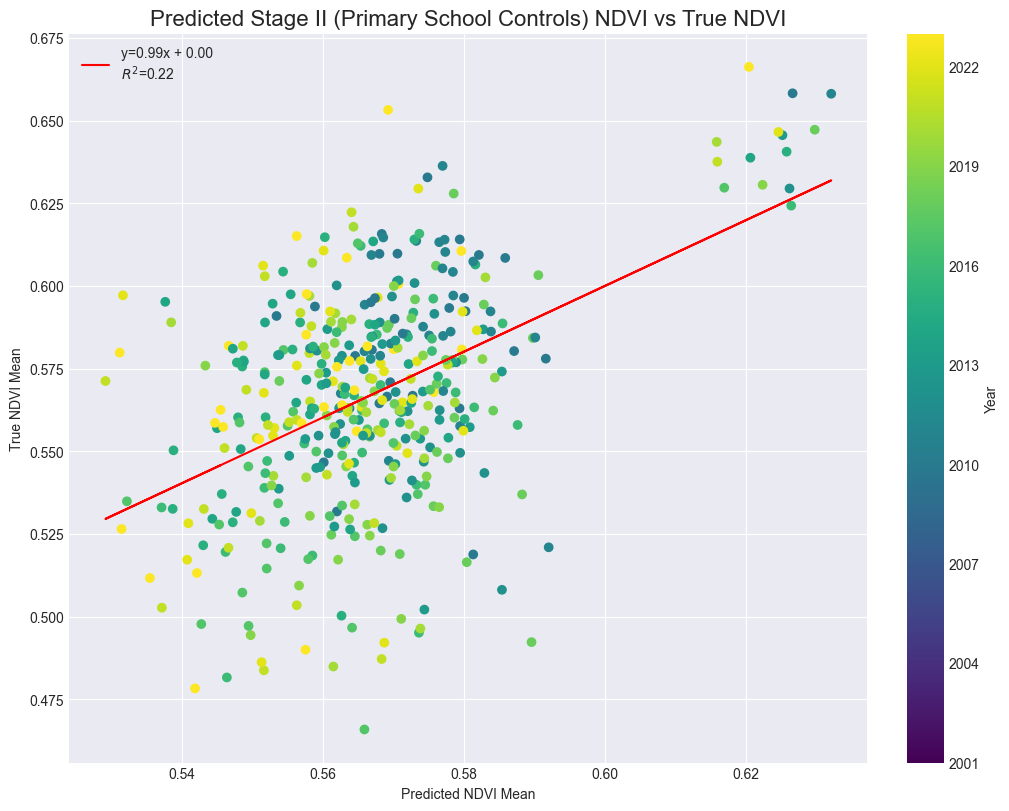}
        \captionof{figure}{Predicted Stage II NDVI (Primary School Controls) vs. True NDVI (2001 - 2023)}
        \label{fig:PredPrimvsTrueNDVI}
\end{figure}

\begin{figure}[h!]
    \centering
        \includegraphics[width=0.7\linewidth]{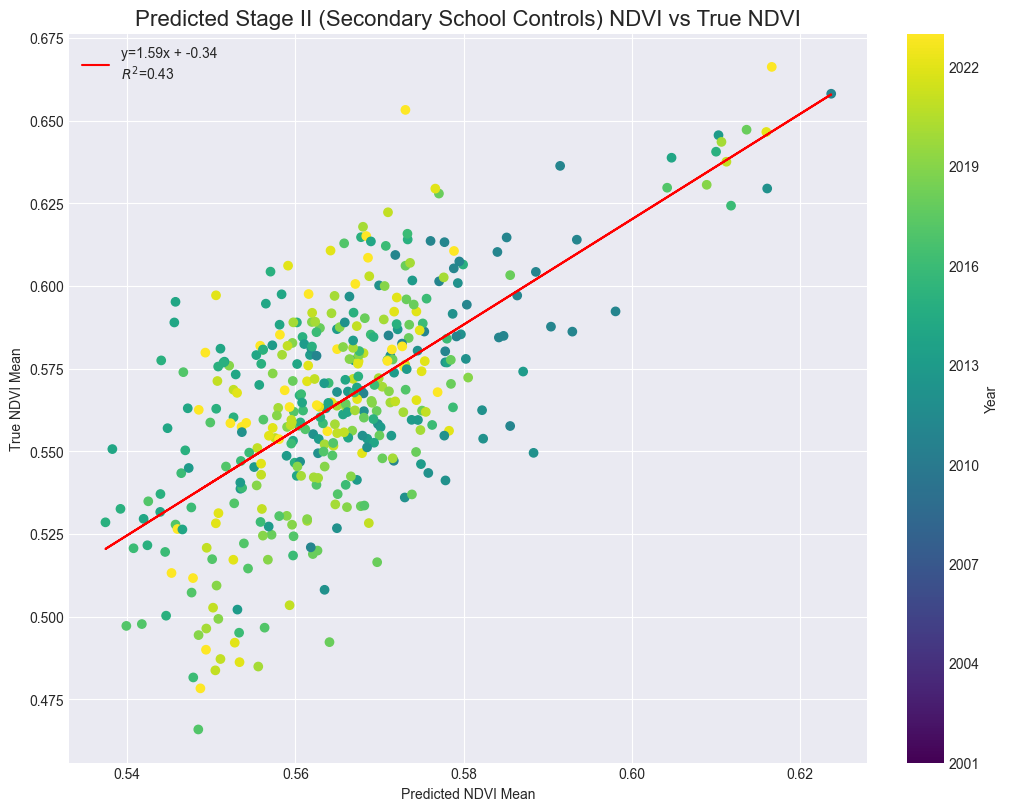}
        \captionof{figure}{Predicted Stage II NDVI (Secondary School Controls) vs. True NDVI (2001 - 2023)}
        \label{fig:PredSecvsTrueNDVI}
\end{figure}

\clearpage
\section{Preliminary Regression Results}
\begin{center}
\scalebox{0.5}{
    \begin{tabularx}{\linewidth}{l*{4}{Y}}
        \toprule
        \multicolumn{4}{l}{\textbf{Panel A: Initial Regression Results for Primary Enrollment}} \\
        \midrule
               & Log Primary Male Enrollment & Log Primary Female Enrollment & Log Primary Total Enrollment \\[0pt]
               &      &        &         \\
    Constant & $8.6925^{***}$ & $9.2603^{***}$   & $9.5558^{***}$      \\
            & (0.1371) & (0.1503) & (0.1416)    \\
            &      &        &         \\
    NDVI &  $-0.9513^{***}$ &  $-0.8443^{***}$   & $-0.8970^{***}$      \\
        & (0.1554) & (0.1703) & (0.1605)    \\
        &      &        &         \\

    Log Primary Total Staff &  0.0649 &  0.0588   & 0.0618      \\
            & (0.0424) & (0.0464) & (0.0438)    \\
            &      &        &         \\

    Log Primary Total Schools &  $0.3895^{***}$ &  $0.4211^{***}$   & $0.4051^{***}$      \\
            & (0.0385) & (0.0422) & (0.0398)    \\
            &      &        &         \\
    Log Primary Total Classrooms &  0.0313 &  -0.0721   & -0.0202      \\
            & (0.0515) & (0.0565) & (0.0532) \\
        \midrule

    Constant & $8.3687^{***}$ & $8.9564^{***}$   & $9.3548^{***}$      \\
            & (0.1497) & (0.1621) & (0.1537)    \\
            &      &        &         \\
    NDVI Lag One Year &  $-0.4183^{**}$ &  $-0.347^{**}$   & $-0.3821^{**}$      \\
        & (0.1660) & (0.1797) & (0.1704)    \\
        &      &        &         \\

    Log Primary Total Staff &  0.0696 &  0.0626   & 0.0661      \\
            & (0.0441) & (0.0478) & (0.0453)    \\
            &      &        &         \\

    Log Primary Total Schools &  $0.4157^{***}$ &  $0.4450^{***}$   & $0.4302^{***}$      \\
            & (0.0398) & (0.0430) & (0.0408)    \\
            &      &        &         \\
    Log Primary Total Classrooms &  0.0128 &  -0.0880   & -0.0373      \\
            & (0.0536) & (0.0580) & (0.0550) \\

         \midrule

    Constant & $8.3695^{***}$ & $8.9071^{***}$   & $9.3360^{***}$      \\
            & (0.1491) & (0.1618) & (0.1691)    \\
            &      &        &         \\
    NDVI Lag Two Years &  $-0.4331^{***}$ &  -0.2740   & $-0.3538^{**}$      \\
        & (0.1645) & (0.1785) & (0.1691)    \\
        &      &        &         \\

    Log Primary Total Staff &  0.0720 &  0.0633   & 0.0676      \\
            & (0.0441) & (0.0478) & (0.0453)    \\
            &      &        &         \\

    Log Primary Total Schools &  $0.4191^{***}$ &  $0.4485^{***}$   & $0.4336^{***}$      \\
            & (0.0396) & (0.0430) & (0.0407)    \\
            &      &        &         \\
    Log Primary Total Classrooms &  0.0072 &  -0.0904   & -0.0414      \\
            & (0.0536) & (0.0582) & (0.0552) \\
        \midrule

    Constant & $8.4558^{***}$ & $9.1591^{***}$   & $9.4996^{***}$      \\
            & (0.1682) & (0.1812) & (0.1722)    \\
            &      &        &         \\
    NDVI Lag Three Years &  $-0.4411^{***}$ &  $-0.5204^{***}$   & $-0.4802^{***}$      \\
        & (0.1645) & (0.1785) & (0.1691)    \\
        &      &        &         \\

    Log Primary Total Staff &  0.0585 &  0.0512   & 0.0548      \\
            & (0.0440) & (0.0474) & (0.0451)    \\
            &      &        &         \\

    Log Primary Total Schools &  $0.4286^{***}$ &  $0.4473^{***}$   & $0.4427^{***}$\\
            & (0.0396) & (0.0426) & (0.0405)    \\
            &      &        &         \\
    Log Primary Total Classrooms &  0.0051 &  $-0.0993^{*}$   & -0.0468      \\
            & (0.0536) & (0.0578) & (0.0549) \\
    \bottomrule
    \end{tabularx}
    \begin{tabularx}{\linewidth}{l*{4}{Y}}
        \toprule
        \multicolumn{4}{l}{\textbf{Panel B: Initial Regression Results for Secondary Enrollment}} \\
        \midrule
               & Log Secondary Male Enrollment & Log Secondary Female Enrollment & Log Secondary Total Enrollment \\[0pt]
               &      &        &         \\
        Constant	 & $4.7676^{***}$	& $4.0889^{***}$	& $5.0873^{***}$ \\
        & (0.2742)	& (0.2881)	& (0.2684) \\
                    &      &        &         \\
        NDVI	 & 0.3607	& $0.4501^{*}$	& $0.4334^{*}$ \\
        & (0.2523)	& (0.2651)	& (0.2469) \\
                    &      &        &         \\
        Log Secondary Total Staff	 & -0.0651	& $-0.3533^{***}$	& $-0.2205^{***}$ \\
        & (0.0483)	& (0.0508)	& (0.0473) \\
                    &      &        &         \\
        Log Secondary Total Schools	 & $0.1811^{***}$	& $0.2486^{***}$	& $0.2181^{***}$ \\
        & (0.0560)	& (0.0588)	& (0.0548) \\
                    &      &        &         \\
        Log Secondary Total Classrooms	 & $0.617^{***}$	& $1.0058^{***}$	& $0.8251^{***}$ \\
        & (0.0565)	& (0.0594)	& (0.0553) \\
        \midrule
        			
        Constant	 & $4.6355^{***}$	& $4.1606^{***}$	& $5.0619^{***}$ \\
        & (0.2973)	& (0.3136)	& (0.2916) \\
                    &      &        &         \\
        NDVI Lag One Year	 & $0.4637^{*}$	& 0.3056	& 0.4034 \\
        & (0.2553)	& (0.2692)	& (0.2504) \\
                    &      &        &         \\
        Log Secondary Total Staff	 & -0.0602	& $-0.3514^{***}$	& $-0.2171^{***}$ \\
        & (0.0484)	& (0.0510)	& (0.0475) \\
                    &      &        &         \\
        Log Secondary Total Schools	 & $0.1858^{***}$	& $0.2623^{***}$	& $0.2282^{***}$ \\
        & (0.0549)	& (0.0580)	& (0.0539) \\
                    &      &        &         \\
        Log Secondary Total Classrooms	 & $0.6204^{***}$	& $0.9967^{***}$	& $0.821^{***}$7 \\
        & (0.0562)	& (0.0593)	& (0.0551) \\
        \midrule
        Constant	 & $5.3443^{***}$	& $4.8426^{***}$	& $5.7666^{***}$ \\
        & (0.2564)	& (0.2687)	& (0.2507) \\
                    &      &        &         \\
        NDVI Lag Two Years	 & $-0.4604^{**}$	& $-0.6255^{**}$	& $-0.5327^{**}$ \\
        & (0.2429)	& (0.2547)	& (0.2376) \\
                    &      &        &         \\
        Log Secondary Total Staff	 & -0.0575	& $-0.3427^{***}$	& $-0.2119^{***}$ \\
        & (0.0485)	& (0.0508)	& (0.0474) \\
                    &      &        &         \\
        Log Secondary Total Schools	 & $0.1926^{***}$	& $0.2623^{***}$	& $0.2323^{***}$ \\
        & (0.0545)	& (0.0572)	& (0.0533) \\
                    &      &        &         \\
        Log Secondary Total Classrooms	 & $0.5839^{***}$	& $0.963^{***}$	& $0.786^{***}$ \\
        & (0.0552)	& (0.0579)	& (0.0540) \\
        	\midrule
        Constant	 & $5.265^{***}$	& $4.8885^{***}$	& $5.7552^{***}$ \\
        & (0.2400)	& (0.2496)	& (0.2339) \\
                    &      &        &         \\
        NDVI Lag Three Years	 & $-0.4065^{*}$	& $-0.8267^{***}$	& $-0.6137^{***}$ \\
        & (0.2399)	& (0.2495)	& (0.2338) \\
                    &      &        &         \\
        Log Secondary Total Staff	 & -0.0629	& $-0.3471^{***}$	& $-0.2166^{***}$ \\
        & (0.0483)	& (0.0503)	& (0.0471) \\
                    &      &        &         \\
        Log Secondary Total Schools	 & $0.1766^{***}$	& $0.2254^{***}$	& $0.2058^{***}$ \\
        & (0.0561)	& (0.0583)	& (0.0546) \\
                    &      &        &         \\
        Log Secondary Total Classrooms	 & $0.6075^{***}$	& $1.0017^{***}$	& $0.8167^{***}$ \\
        & (0.0551)	& (0.0573)	& (0.0537) \\
        \bottomrule
    \end{tabularx}}
\end{center}
\begin{center}
    \textit{Notes: All regressions include district and year fixed effects.\\ Standard errors are in parentheses.\\ $^{*}$ denotes significance at 10\% level; $^{**}$ denotes significance at 5\% level; $^{***}$ denotes significance at 1\% level}
\end{center}

\end{appendices}

\clearpage
\bibliographystyle{plain}

\end{document}